\documentclass[prd,letterpaper,twocolumn,showpacs,showkeys,lengthcheck]{revtex4}
\usepackage{graphicx}% Include figure files
\usepackage{dcolumn}% Align table columns on decimal point
\usepackage{bm}% bold math
\usepackage{comment}
\begin{document}
%%%%% Personal Macros %%%%%%%%%%%%%%%%%%%
% Journal macro
\def\Journal#1#2#3#4{{#1} {\bf #2}, #3 (#4)}
% Journal names
\def\AHEP{Advances in High Energy Physics.} 	
\def\AACA{Advances in Applied Clifford Algebras.} 	
\def\ARNPS{Annu. Rev. Nucl. Part. Sci.} 
\def\AandA{Astron. Astrophys.} 
\def\ANP{Ann. Phys.}
\def\APJ{Astrophys. J.}
\def\APJS{Astrophys. J. Suppl}
\def\COMR{Comptes Rendues}
\def\CQG{Class. Quantum Grav.}
\def\CPC{Chin. Phys. C}
\def\EPJC{Eur. Phys. J. C}
\def\EPL{EPL}
\def\IJMPA{Int. J. Mod. Phys. A}
\def\IJMPE{Int. J. Mod. Phys. E}
\def\JCAP{J. Cosmol. Astropart. Phys.}
\def\JHEP{J. High Energy Phys.}
\def\JETPL{JETP. Lett.}
\def\JETPUSSR{JETP (USSR)}
\def\JPG{J. Phys. G} 
\def\JPCS{J. Phys. Conf. Ser.} 
\def\JPGNP{J. Phys. G: Nucl. Part. Phys.} 
\def\JNP{J. of Number Theor.} 
\def\MPLA{Mod. Phys. Lett. A}
\def\NIMA{Nucl. Instrum. Meth. A.}
\def\NATU{Nature}
\def\NCA{Nuovo Cimento}
\def\NJP{New. J. Phys.}
\def\NPB{Nucl. Phys. B}
\def\NPBOLD{Nucl. Phys.}
\def\NPBSUPPL{Nucl. Phys. B. Proc. Suppl.}
\def\PL{Phys. Lett.}
\def\PLB{{Phys. Lett.} B}
\def\PMCA{PMC Phys. A}
\def\PREP{Phys. Rep.}
\def\PPNP{Prog. Part. Nucl. Phys.}
\def\PLBOLD{Phys. Lett.}
\def\PAN{Phys. Atom. Nucl.}
\def\PRL{Phys. Rev. Lett.}
\def\PRD{Phys. Rev. D}
\def\PRC{Phys. Rev. C}
\def\PR{Phys. Rev.}
\def\PTP{Prog. Theor. Phys.}
\def\PTEP{Prog. Theor. Exp. Phys.}
\def\RMP{Rev. Mod. Phys.}
\def\RPP{Rep. Prog. Phys.}
\def\SJNP{Sov. J. Nucl. Phys.}
\def\SCIENCE{Science}
\def\SPJETP{Sov. Phys. JETP.}
\def\TNYAS{Trans. New York Acad. Sci.}
\def\ZETP{Zh. Eksp. Teor. Piz.}
\def\ZFPH{Z. fur Physik}
\def\ZPC{Z. Phys. C}
%\preprint{APS/123-QED}
\title{Primitive Pythagorean triples and neutrino mixing}
\author{Yuta Hyodo}
\email[Corresponding author:~]{2CTAD004@mail.u-tokai.ac.jp}
\affiliation{Graduate School of Science and Technology, Tokai University, 
4-1-1 Kitakaname, Hiratsuka, Kanagawa 259-1292, Japan}
\affiliation{Micro/Nano Technology Center, Tokai University,
4-1-1 Kitakaname, Hiratsuka, Kanagawa 259-1292, Japan}
\author{Teruyuki Kitabayashi}
\email{teruyuki@tokai-u.jp}
\affiliation{Department of Physics, Tokai University,\\
4-1-1 Kitakaname, Hiratsuka, Kanagawa 259-1292, Japan}
\date{\today}% It is always \today, today,
\begin{abstract}
The primitive Pythagorean triples are the three natural numbers $(a, b, c)$ that satisfy $c^2=a^2+b^2$ in a right triangle. We constructed a neutrino mixing models related to primitive Pythagorean triples that satisfy the observed values within the $3\sigma$ region for the reactor, solar, and atmospheric neutrino mixing angles, as well as the Dirac CP phase.
\end{abstract}
\maketitle
%\tableofcontents
\section{Introduction\label{section:introduction}}
\begin{figure}[h]
\hspace{4cm}
\includegraphics[keepaspectratio, scale=0.5]{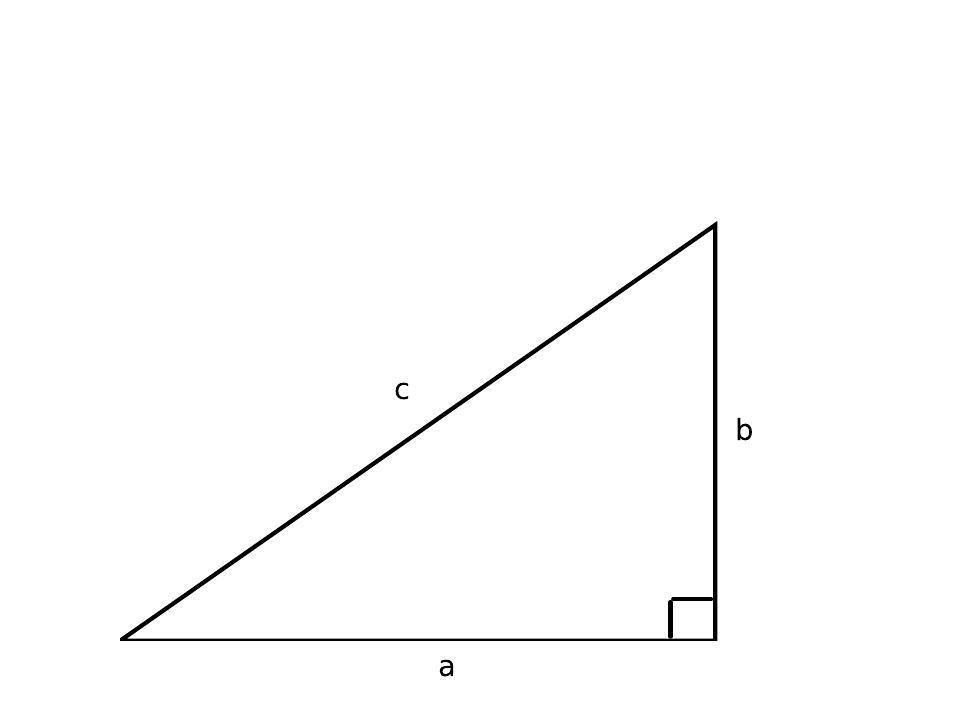}
 \caption{Right triangle.}
 \label{fig:triangle}
  \end{figure}

The primitive Pythagorean triples\cite{Williams2017,MartensarXiv2112,Kocik2007,SchmelzerarXiv2021,YekutieliarXiv2021,Cha2018,DeriyarXiv2023,PricearXiv2008,AlperinarXiv2000} are the three natural numbers $(a, b, c)$ that satisfy $c^2=a^2+b^2$ in the right triangle (Fig. \ref{fig:triangle}). The most famous primitive Pythagorean triple is the combination $(3, 4, 5)$.  The internal angles of a triangle with the combination $(3, 4, 5)$ are $(36.87^\circ,53.13^\circ,90.00^\circ)$. 
%--------------------------------------------------------------------
%--------------------------------------------------------------------
The two internal angles $(36.87^\circ,53.13^\circ)$ closely approximate the upper limit of the $3 \sigma$ region for the so-called solar neutrino mixing angle, $35.74^\circ$, and the upper limit of the $3 \sigma$ region for the so-called atmospheric neutrino mixing angle, $51.0^\circ~\left(51.5^\circ\right)$, in the case of normal mass ordering (NO) [inverted mass ordering (IO)]. In this paper, we constructed a neutrino mixing model for the solar and atmospheric mixing angles with primitive Pythagorean triples.

The rest of the paper is organized as follows. In Section. \ref{section:PMNS}, we review the Pontecorvo-Maki-Nakagawa-Sakata (PMNS) mixing matrix and observed values to compare the mixing model with observed values. In Section. \ref{section:PT} describes the construction of a neutrino mixing model related to primitive Pythagorean triples. In Section. \ref{section:NC} presents the numerical calculation results of the neutrino mixing model constructed in Section. \ref{section:PT}. In Section. \ref{section:Z2} describes the investigation of if the neutrino mass matrix from the proposed neutrino mixing model is invariant under $Z_2$ symmetry. Section. \ref{section:summary} presents the summary.

%=================================================
%=================================================
%%----------------------------------------------------------------------------------
\section{PMNS Matrix\label{section:PMNS}}
%%----------------------------------------------------------------------------------
We review the PMNS mixing matrix\cite{Pontecorvo1957,Pontecorvo1958,Maki1962PTP,PDG} and observed values.

PMNS matrix is written as
\begin{widetext}
\begin{eqnarray}
U  
&=&\left ( 
\begin{array}{ccc}
U_{e1}&U_{e2}&U_{e3}\\
U_{\mu1}&U_{\mu2}&U_{\mu3}\\
U_{\tau1}&U_{\tau2}&U_{\tau3}
\end{array}
\right)\nonumber\\
&=&
\left ( 
\begin{array}{ccc}
c_{12}c_{13} & s_{12}c_{13} & s_{13} e^{-i\delta} \\
- s_{12}c_{23} - c_{12}s_{23}s_{13} e^{i\delta} & c_{12}c_{23} - s_{12}s_{23}s_{13}e^{i\delta} & s_{23}c_{13} \\
s_{12}s_{23} - c_{12}c_{23}s_{13}e^{i\delta} & - c_{12}s_{23} - s_{12}c_{23}s_{13}e^{i\delta} & c_{23}c_{13} \\
\end{array}
\right),
\end{eqnarray}
\end{widetext}
where, $c_{ij}=\cos\theta_{ij}$, $s_{ij}=\sin\theta_{ij}$ ($i,j$=1,2,3), $\theta_{ij}$ is a mixing angle, and $\delta$ is the Dirac CP phase. The sines and cosines of the three neutrino mixing angles of the PMNS matrix, $U$, are given by
\begin{eqnarray}
s_{12}^2&=&\frac{|U_{e2}|^2}{1-|U_{e3}|^2},~~ s_{23}^2=\frac{|U_{\mu3}|^2}{1-|U_{e3}|^2},~~s_{13}^2=|U_{e3}|^2,\nonumber \\
c_{12}^2&=&\frac{|U_{e1}|^2}{1-|U_{e3}|^2},~~c_{23}^2=\frac{|U_{\tau3}|^2}{1-|U_{e3}|^2}.
\label{Eq:mixing_angle_PMNS}
\end{eqnarray}

The Jarlskog rephasing invariant\cite{Jarlskog},
\begin{eqnarray}
J=\rm{Im}(U_{e1}U_{e2}^{\ast}U_{\mu1}^{\ast}U_{\mu2})=s_{12}s_{23}s_{13}c_{12}c_{23}c_{13}^2\sin{\delta}
\label{Eq:Jarlskog}
\end{eqnarray}
can be used to calculate $\delta$.

A global analysis of current data reveals the following the best-fit values of  the neutrino mixing angles in the case of NO, $m_1<m_2<m_3$, where $m_1,m_2,$ and $m_3$ are the neutrino mass eigenvalues, \cite{NuFit}:
\begin{eqnarray} 
s_{12}^2&=& 0.303^{+0.012}_{-0.012} \quad (0.270 \sim 0.341), \nonumber \\
s_{23}^2&=& 0.451^{+0.019}_{-0.016} \quad (0.408 \sim 0.603), \nonumber \\
s_{13}^2&=& 0.02225^{+0.00056}_{-0.00059} \quad (0.02052 \sim 0.02398), \nonumber \\
\delta/^\circ &=& 232^{+36}_{-26} \quad (144 \sim 350), 
\label{Eq:neutrino_observation_NO}
\end{eqnarray}
where the $\pm$ denotes the $1 \sigma$ region and the parentheses denote the $3 \sigma$ region. For the IO, $m_3 < m_1<m_2$, we have
\begin{eqnarray} 
s_{12}^2&=& 0.303^{+0.012}_{-0.012} \quad (0.270 \sim 0.341), \nonumber \\
s_{23}^2 &=& 0.569^{+0.016}_{-0.021} \quad (0.412 \sim 0.613), \nonumber \\
s_{13}^2&=& 0.02223^{+0.00058}_{-0.00058} \quad (0.02048 \sim 0.02416), \nonumber \\
\delta/^\circ &=& 276^{+22}_{-29} \quad (194  \sim 344).
\label{Eq:neutrino_observation_IO}
\end{eqnarray}
%
%=================================================
%=================================================
%%----------------------------------------------------------------------------------
\section{Primitive primitive Pythagorean triples and neutrino mixing model\label{section:PT}}
%%----------------------------------------------------------------------------------
%%----------------------------------------------------------------------------------
\subsection{Primitive primitive Pythagorean triples \label{subsection:PT}}
%%----------------------------------------------------------------------------------
Primitive Pythagorean triples are the combinations $(a, b, c)$ of three natural numbers that satisfy $a^2+b^2=c^2$. The relationship among $a, b, c$ is described by
\begin{eqnarray} 
a<b<c.
\end{eqnarray}
$a,b,c$ can be expressed in terms of two natural numbers, $m$ and $n$, using either
\begin{eqnarray} 
\left(a,b,c\right)=\left(m^2-n^2, 2mn, m^2+n^2\right)
\label{Eq:Primitive_pythagorean_triples1}
\end{eqnarray}
or
\begin{eqnarray} 
\left(a,b,c\right)=\left(2mn, m^2-n^2, m^2+n^2\right)
\label{Eq:Primitive_pythagorean_triples2}
\end{eqnarray}
\cite{MartensarXiv2112,Kocik2007,SchmelzerarXiv2021}. In the case of Eq. (\ref{Eq:Primitive_pythagorean_triples1}) (Eq. (\ref{Eq:Primitive_pythagorean_triples2})), $a$ is an odd (even) number, and $b$ is an even (odd) number.

The relationship between $m$ and $n$ is as follows\cite{MartensarXiv2112,Kocik2007,SchmelzerarXiv2021}:
\begin{itemize}
\item $m>n$
\item $m$ and $n$ are natural numbers.
\item The parities of $m$ and $n$. (one is an odd number; and the other is an even number)
\item $m$ and $n$ are coprime.
\end{itemize}
%%----------------------------------------------------------------------------------
\subsection{Relationship between the solar mixing angle, atmospheric mixing angle, and primitive Pythagorean triples\label{subsection:PTmixing}}
%%----------------------------------------------------------------------------------
The mixing matrix predicting a reactor mixing angle ($\theta_{13}$) of $0$ can be written as
\begin{eqnarray}
U=R_{23}R_{12}.
\label{Eq:Uv}
\end{eqnarray}
$R_{23}$ $(R_{12})$ is a rotation matrix for the 2-3 (1-2) plane\cite{PDG}; it is denoted by
\begin{eqnarray}
R_{23}=\left(
\begin{array}{ccc}
1 & 0  & 0 \\
0 &  \cos{\theta_{23}}  & \sin{\theta_{23}}   \\
0 &  -\sin{\theta_{23}}  & \cos{\theta_{23}} 
\end{array}
\right),\\
R_{12}=\left(
\begin{array}{ccc}
\cos{\theta_{12}} & \sin{\theta_{12}}   & 0 \\
-\sin{\theta_{12}}&  \cos{\theta_{12}}  & 0  \\
0 &  0  &1
\end{array}
\right).
\label{Eq:Rotation_Matrix}
\end{eqnarray}
The relationship between primitive Pythagorean triples and neutrino mixing angles, $\theta_{12}$ and $\theta_{23}$, is as follows: 
\begin{eqnarray} 
s_{12}=\frac{a}{c},~~s_{23}=\frac{b}{c}.
\label{Eq:PT_12_23_1}
\end{eqnarray}

With the relationship provided in Eq. (\ref{Eq:PT_12_23_1}), the mixing matrix predicting a reactor mixing angle of $0$ can be written as
\begin{eqnarray}
U_{\rm{PM}}=\left(
\begin{array}{ccc}
\frac{b}{c} & \frac{a}{c}  & 0 \\
-\frac{a^2}{c^2} &  \frac{ab}{c^2}  & \frac{b}{c}   \\
-\frac{ab}{c^2} &  -\frac{b^2}{c^2}   & \frac{a}{c}
\end{array}
\right).
\label{Eq:PM1}
\end{eqnarray}
We call this mixing matrix primitive Pythagorean triple neutrino mixing (PM).

The mixing matrix predicts the vanishing reactor mixing angle. Next, we performed modifications to improve the prediction of the reactor mixing angle.
%%----------------------------------------------------------------------------------
\subsection{Improved reproducibility of reactor mixing angle\label{subsection:reactor}}
%%----------------------------------------------------------------------------------
To improve the reproducibility of the reactor mixing angle, we employed the method of changing from tri-bimaximal (TBM) mixing to trimaximal (TM) mixing\cite{Xing2007PLB,Harrison2006PRD}. Using this method, the first column (second column) of the $U_{\rm{PM}}$ remains unchanged, whereas the second and third columns (first and third) are adjusted.

%%----------------------------------------------------------------------------------
\subsubsection{Modifications of the second and third columns\label{section:reactorC1}}
%%----------------------------------------------------------------------------------
As shown in Ref. \cite{Xing2007PLB}, the first column of the $U_{\rm{PM}}$ remains unchanged, whereas the second and third columns are adjusted as follows:
\begin{widetext}
\begin{eqnarray}
\left(U_{\rm{PM}}\right)_{\rm{C1}}&=&
U_{\rm{PM}}U_{23}\nonumber\\
&=&\left(
\begin{array}{ccc}
\frac{b}{c} & \frac{a}{c}\cos{\theta}  & \frac{a}{c}\sin{\theta}e^{-i\phi} \\
-\frac{a^2}{c^2} & \frac{b}{c^2} \left(a\cos{\theta}-c\sin{\theta}e^{i\phi} \right)  & \frac{b}{c^2} \left(c\cos{\theta}+a\sin{\theta}e^{-i\phi} \right)\\
\frac{ab}{c^2} &    -\frac{1}{c^2}\left(b^2\cos{\theta}+ac\sin{\theta}e^{i\phi}\right) &\frac{1}{c^2}\left( ac \cos{\theta}-b^2\sin{\theta}e^{-i\phi}\right) 
\end{array}
\right),
\label{Eq:PT1C1}
\end{eqnarray}
\end{widetext}
where
\begin{eqnarray}
U_{23}=\left(
\begin{array}{ccc}
1&0&0\\
0&\cos{\theta}&\sin{\theta}e^{-i\phi}\\
0&-\sin{\theta}e^{i\phi}&\cos{\theta}
\end{array}
\right),
\label{Eq:U23}
\end{eqnarray}
$\theta$ denotes a rotation angle, and $\phi$ denotes a phase parameter. 

From Eq. (\ref{Eq:mixing_angle_PMNS}), the neutrino mixing angles of $\left(U_{\rm{PM}}\right)_{\rm{C1}}$ are as follows:
\begin{eqnarray} 
s_{12}^2&=& \frac{a^2\cos^2{\theta}}{c^2-a^2\sin^2{\theta}},\nonumber \\
s_{23}^2&=& \frac{b^2c^2\cos^2{\theta}+ab^2\left(a\sin^2{\theta}+c\sin{2\theta}\cos{\phi}\right)}{c^4-a^2c^2\sin^2{\theta}}, \nonumber \\
s_{13}^2&=& \frac{a^2}{c^2}\sin^2{\theta}.
\label{Eq:PT1C1_mixingangle}
\end{eqnarray}
In term of $s_{13}^2$, detailing $s_{12}^2$ and $s_{23}^2$ leads to
\begin{eqnarray} 
s_{12}^2&=& \frac{a^2}{c^2}\times\frac{\cos^2{\theta}}{1-s_{13}^2},\nonumber \\
s_{23}^2&=&\frac{b^2}{c^3}\times\frac{a\sin{2\theta}\cos{\phi}+c\left(\cos^2{2\theta}+s_{13}^2\right)}{1-s_{13}^2}.
\end{eqnarray}

From Eq. (\ref{Eq:Jarlskog}), the Dirac CP phase, $\delta$, is written as follows:
\begin{eqnarray} 
&&\tan{\delta}=\frac{a(c^2+b^2+a^2\cos{2\theta})\sin{\phi}}{a\{a^2+(c^2+b^2)\cos{2\theta}\}\cos{\phi}+c(a-b)(a+b)\sin{2\theta}},\nonumber\\
&&(a=m^2-n^2,b=2mn,c=m^2+n^2),
\label{Eq:PT1C1_tandelta}
\end{eqnarray}
and
\begin{eqnarray} 
&&\tan{\delta}=\frac{a(2c^2-a^2+a^2\cos{2\theta})\sin{\phi}}{-a\{a^2+(2c^2-a^2)\cos{2\theta}\}\cos{\phi}+c(b-a)(b+a)\sin{2\theta}},\nonumber\\
&&(a=2mn,b=m^2-n^2,c=m^2+n^2).
\label{Eq:PT2C1_tandelta}
\end{eqnarray}
%
%--------------------------------------------------------------------
\begin{figure*}[t]
\includegraphics[keepaspectratio, scale=0.7]{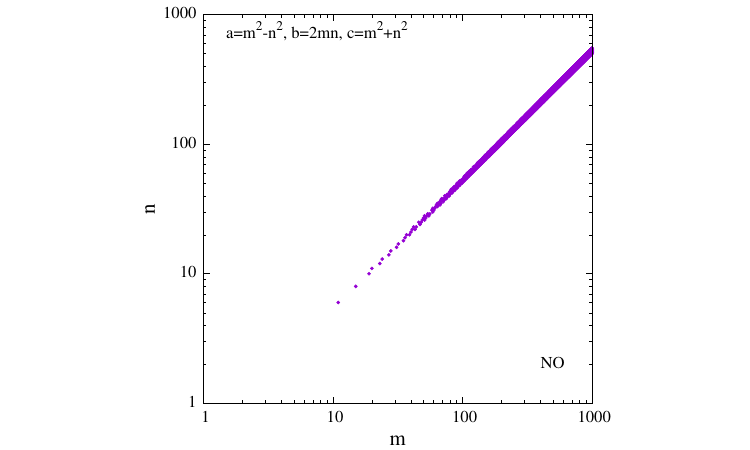}
\includegraphics[keepaspectratio, scale=0.7]{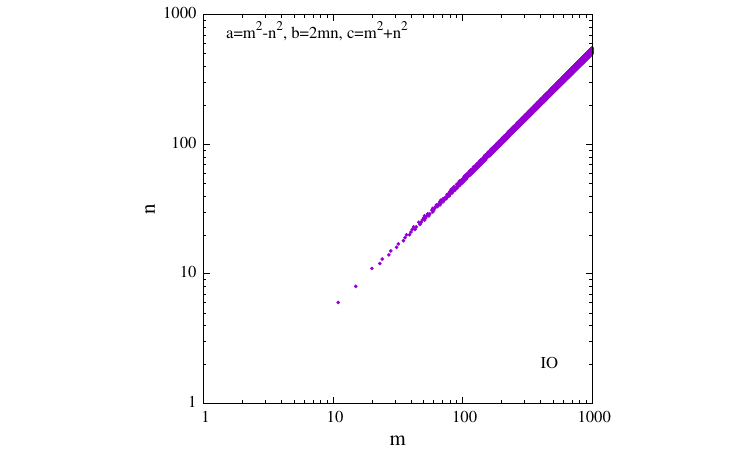}
 \caption{Dependence of $m$ on $n$ of $\left(U_{\rm{PM}}\right)_{\rm{C1}}$ in the case of  $\left(a,b,c\right)=\left(m^2-n^2, 2mn, m^2+n^2\right)$. $\theta$, and $\phi$ varied within the ranges of $0$ to $\frac{\pi}{2}$ and $0$ to $2\pi$, respectively.}
 \label{fig:m_n_PM1C1}
  \end{figure*}
%--------------------------------------------------------------------
%--------------------------------------------------------------------
\begin{figure*}[t]
\includegraphics[keepaspectratio, scale=0.7]{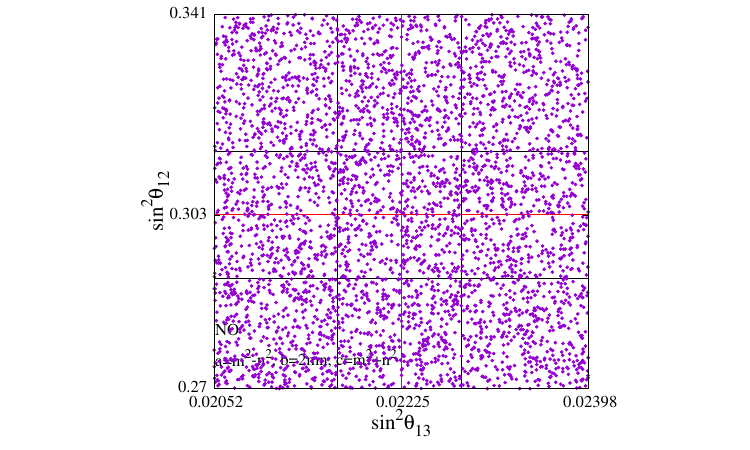}
\includegraphics[keepaspectratio, scale=0.7]{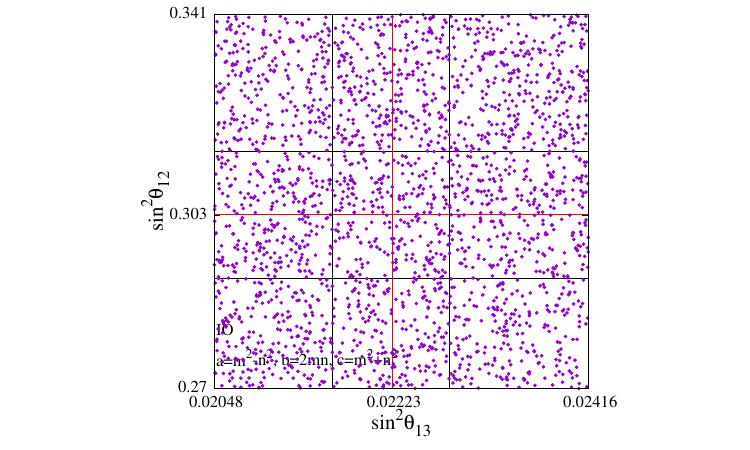}
\includegraphics[keepaspectratio, scale=0.7]{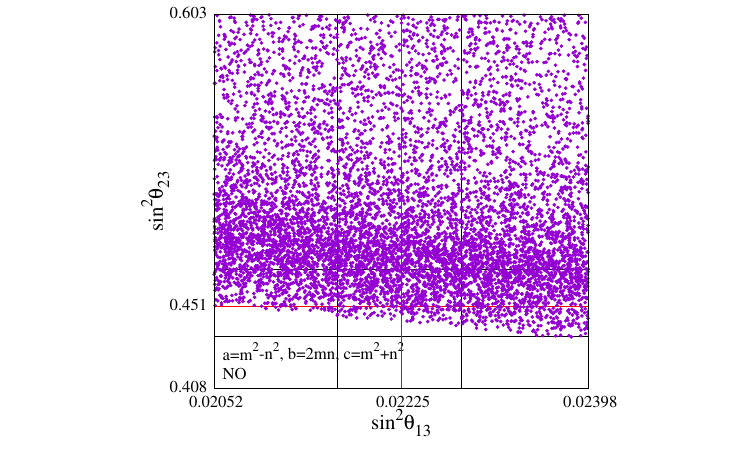}
\includegraphics[keepaspectratio, scale=0.7]{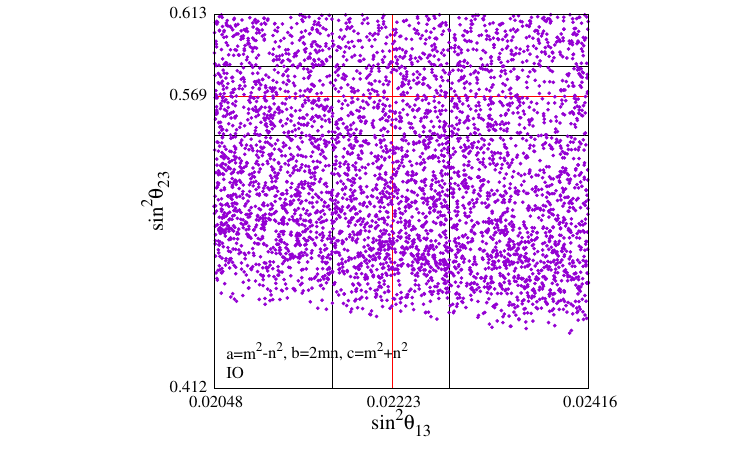}
 \caption{Dependence of $s_{12}^2$ and $s_{23}^2$ on $s_{13}^2$ of $\left(U_{\rm{PM}}\right)_{\rm{C1}}$ in the case of  $\left(a,b,c\right)=\left(m^2-n^2, 2mn, m^2+n^2\right)$. $\theta$, $\phi$ and $m$ and $n$ varied within the ranges of $0 \sim \frac{\pi}{2}$, the range of $\phi$ is varied within $0 \sim 2\pi$ and the range of $m$ and $n$ is varied within $0$ to $\frac{\pi}{2}$ and $0$ to $2\pi$, and $0\sim1000$, respectively.}
 \label{fig:12_13_23_PM1C1}
  \end{figure*}
%--------------------------------------------------------------------
%--------------------------------------------------------------------
\begin{figure*}[t]
\includegraphics[keepaspectratio, scale=0.7]{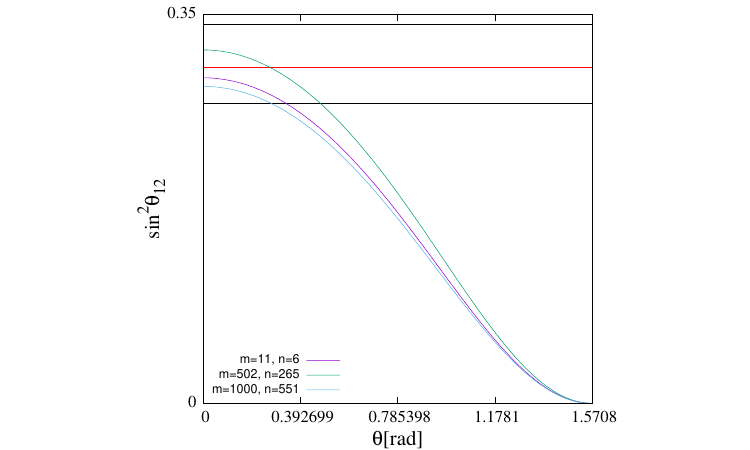}\\
\includegraphics[keepaspectratio, scale=0.7]{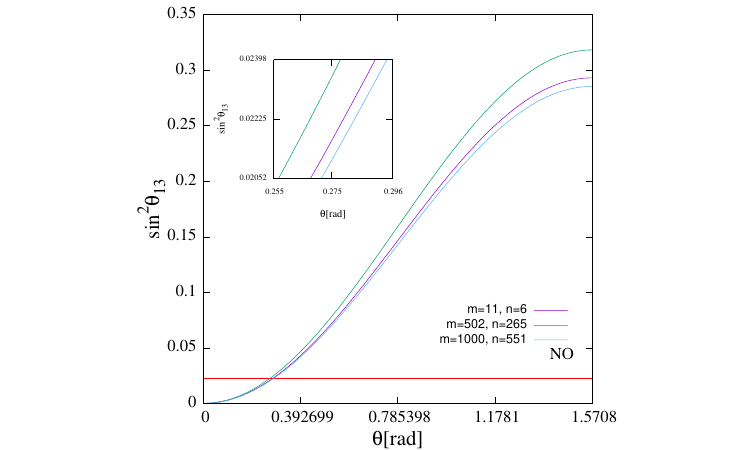}
\includegraphics[keepaspectratio, scale=0.7]{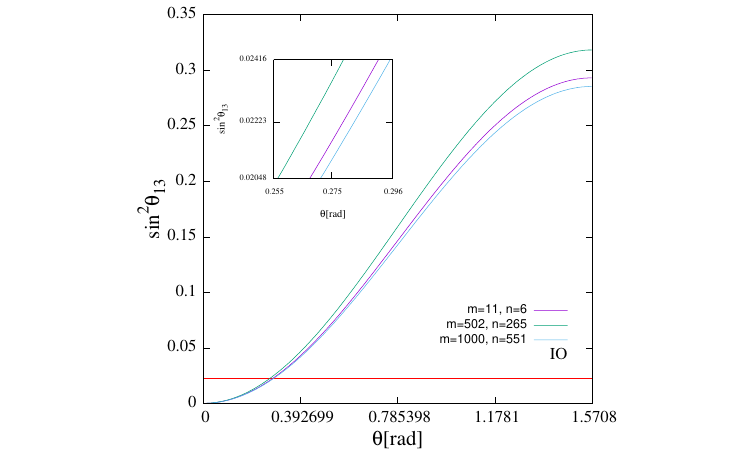}
 \caption{Dependence of $s_{12}^2$ and $s_{13}^2$ on $\theta$ in the cases of $m=11,n=6$, $m=502, n=265$ and $m=1000,n=551$.}
 \label{fig:12_13_t_PM1C1}
  \end{figure*}
%--------------------------------------------------------------------
%--------------------------------------------------------------------
\begin{figure*}[th]
\includegraphics[keepaspectratio, scale=0.7]{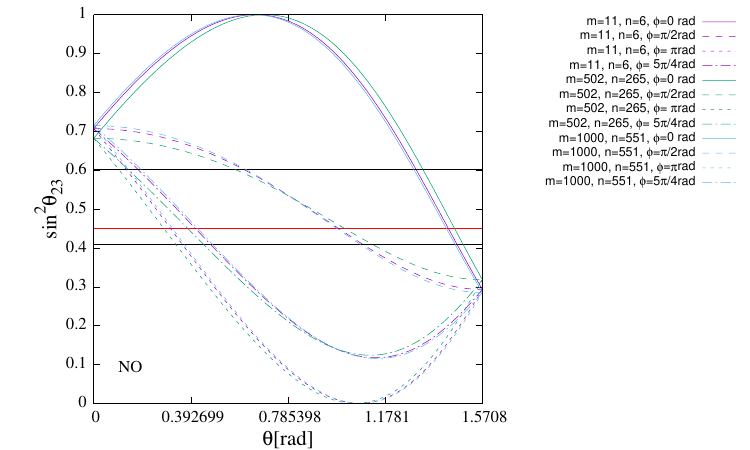}
\includegraphics[keepaspectratio, scale=0.7]{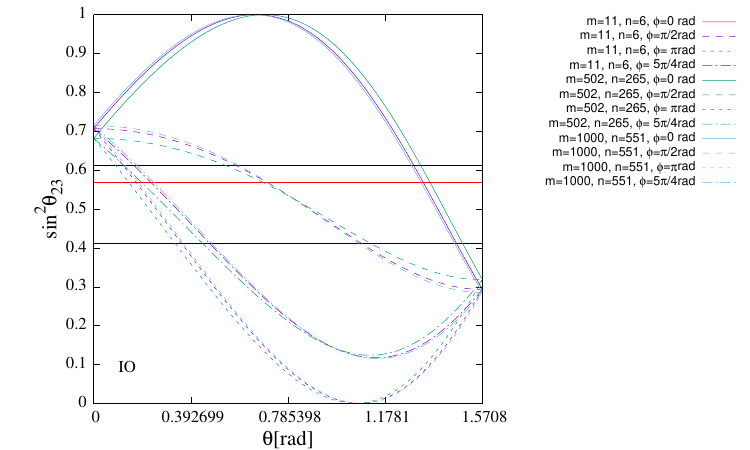}
 \caption{Dependence of $s_{23}^2$ on $\theta$ in the cases of $m=11,n=6$, $m=502, n=265$ and $m=1000,n=551$. We selected $0$,$\frac{\pi}{2}$,$\pi$,$\frac{5\pi}{4}$ for $\phi$.}
 \label{fig:23_t_PM1C1}
  \end{figure*}
%--------------------------------------------------------------------
%--------------------------------------------------------------------
\begin{figure*}[th]
\includegraphics[keepaspectratio, scale=0.7]{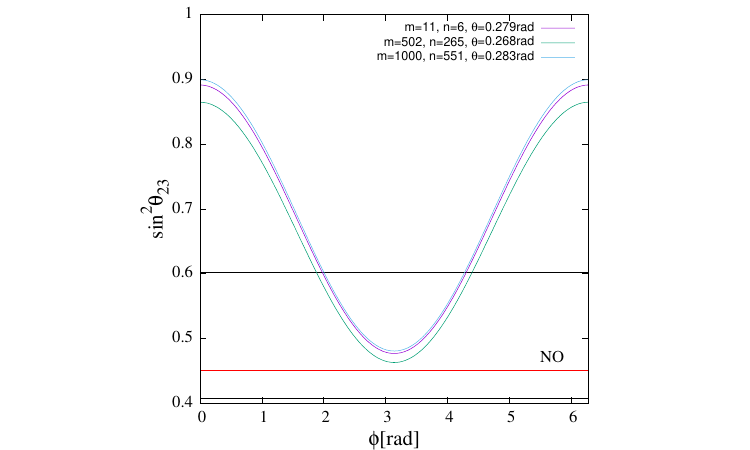}
\includegraphics[keepaspectratio, scale=0.7]{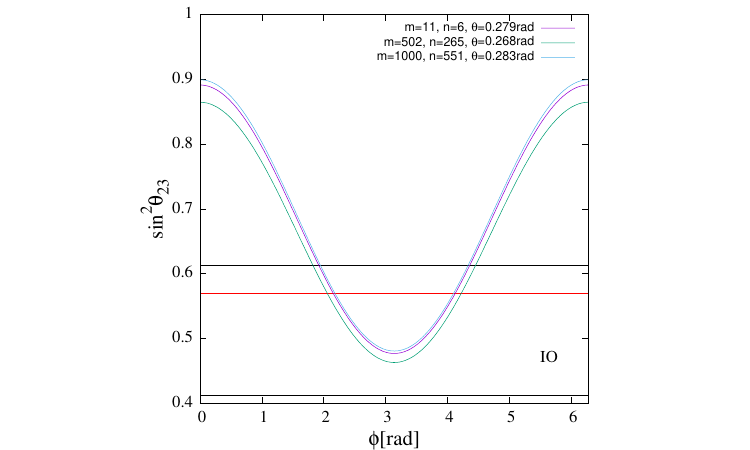}
 \caption{Dependence of $s_{23}^2$ on $\phi$ in the cases of $m=11,n=6$, $m=502, n=265$ and $m=1000,n=551$. We selected $\theta$ such that $s_{13}^2$ is approximately $0.022$.}
 \label{fig:23_p_PM1C1}
  \end{figure*}
%--------------------------------------------------------------------
%--------------------------------------------------------------------
\begin{figure*}[th]
\includegraphics[keepaspectratio, scale=0.7]{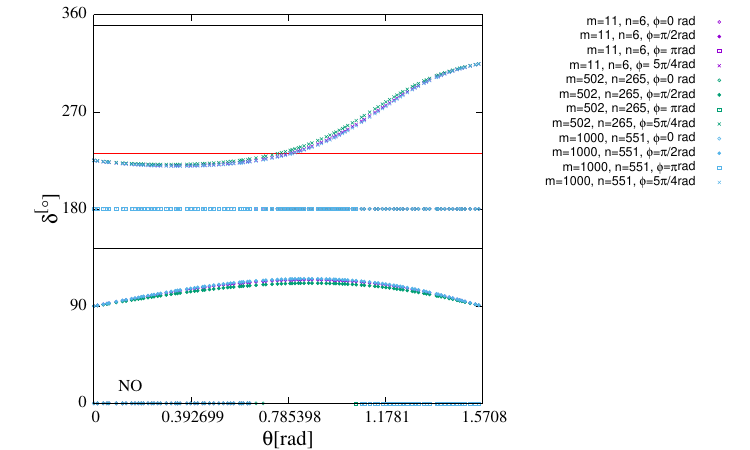}
\includegraphics[keepaspectratio, scale=0.7]{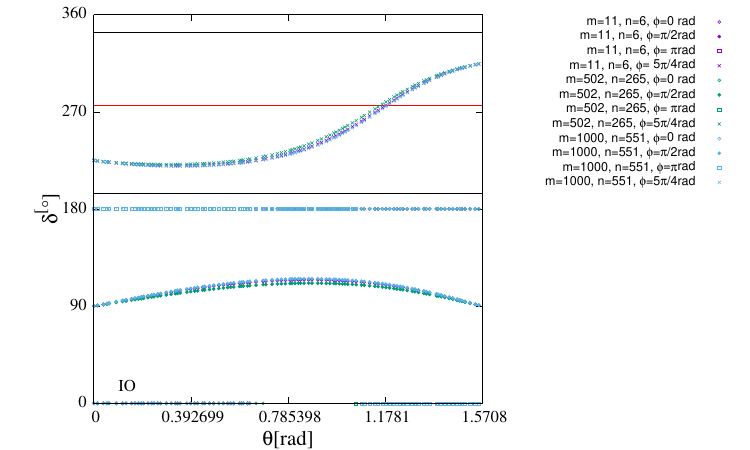}
 \caption{Dependence of Dirac CP phase $\delta$ on $\theta$ in the cases of $m=11,n=6$, $m=502, n=265$ and $m=1000,n=551$. We selected $0$,$\frac{\pi}{2}$,$\pi$,$\frac{5\pi}{4}$ for $\phi$.}
 \label{fig:d_t_PM1C1}
  \end{figure*}
%--------------------------------------------------------------------
%--------------------------------------------------------------------
\begin{figure*}[th]
\includegraphics[keepaspectratio, scale=0.7]{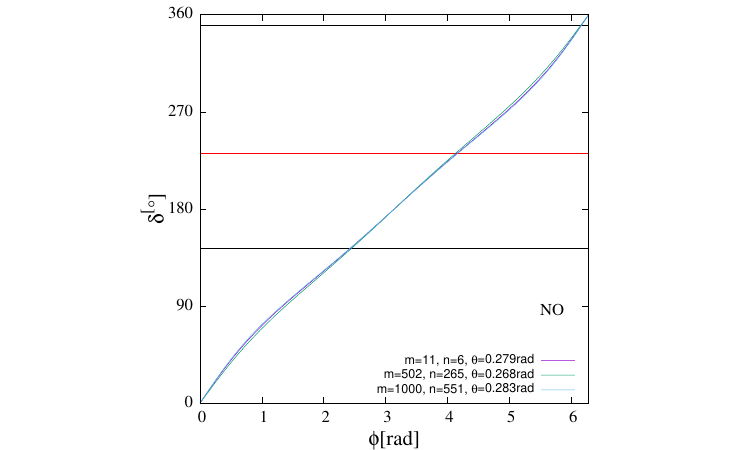}
\includegraphics[keepaspectratio, scale=0.7]{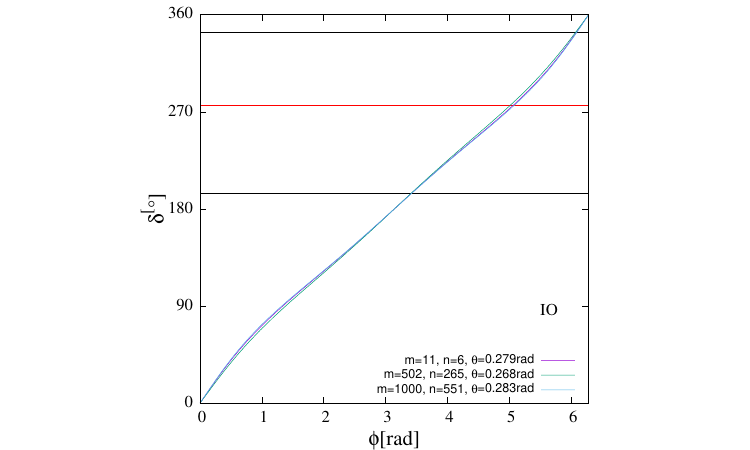}
 \caption{Dependence of Dirac CP phase $\delta$ on $\phi$ in the cases of $m=11,n=6$, $m=502, n=265$ and $m=1000,n=551$. We selected $\theta$ such that $s_{13}^2$ is approximately $0.022$.}
 \label{fig:d_p_PM1C1}
  \end{figure*}
%--------------------------------------------------------------------
%%----------------------------------------------------------------------------------
\subsubsection{Modifications first and third columns\label{section:reactorC2}}
%%----------------------------------------------------------------------------------
As shown in Ref. \cite{Harrison2006PRD}, the second column of the $U_{\rm{PM}}$ remains unchanged, whereas the first and third columns are adjusted as follows:
\begin{widetext}
\begin{eqnarray}
\left(U_{\rm{PM}}\right)_{\rm{C2}}&=&
U_{\rm{PM}}U_{13}\nonumber\\
&=&\left(
\begin{array}{ccc}
 \frac{b}{c}\cos{\theta} &\frac{a}{c}&\frac{b}{c}\sin{\theta}e^{-i\phi}\\
-\frac{1}{c^2}\left(a^2\cos{\theta}+bc\sin{\theta}e^{i\phi}\right)&\frac{ab}{c^2}&\frac{1}{c^2}\left(bc\cos{\theta}-a^2\sin{\theta}e^{-i\phi}\right)\\
\frac{a}{c^2} \left(b\cos{\theta}-c\sin{\theta}e^{i\phi} \right) &-\frac{b^2}{c^2}&\frac{a}{c^2} \left(c\cos{\theta}+b\sin{\theta}e^{-i\phi} \right)
\end{array}
\right),
\label{Eq:PT1C2}
\end{eqnarray}
\end{widetext}
where
\begin{eqnarray}
U_{13}=\left(
\begin{array}{ccc}
\cos{\theta}&0&\sin{\theta}e^{-i\phi}\\
0&1&0\\
-\sin{\theta}e^{i\phi}&0&\cos{\theta}
\end{array}
\right),
\label{Eq:U13}
\end{eqnarray}
where $\theta$ denotes a rotation angle, and $\phi$ denotes a phase parameter. 

From Eq. (\ref{Eq:mixing_angle_PMNS}), the neutrino mixing angles of $\left(U_{\rm{PM}}\right)_{\rm{C2}}$ are as follows:
\begin{eqnarray} 
s_{12}^2&=& \frac{a^2}{c^2-b^2\sin^2{\theta}},\nonumber \\
s_{23}^2&=& \frac{b^2c^2\cos^2{\theta}-a^2bc\sin{2\theta}\cos{\phi}+a^4\sin^2{\theta}}{c^4-b^2c^2\sin^2{\theta}}, \nonumber \\
s_{13}^2&=& \frac{b^2}{c^2}\sin^2{\theta}.
\label{Eq:PT1C2_mixingangle}
\end{eqnarray}
In term of $s_{13}^2$, detailing $s_{12}^2$ and $s_{23}^2$ leads to
\begin{eqnarray} 
s_{12}^2&=& \frac{a^2}{c^2}\times\frac{1}{1-s_{13}^2},\nonumber \\
s_{23}^2&=&\frac{b^2c^2\cos^2{\theta}-a^2bc\sin{2\theta}\cos{\phi}+a^4\sin^2{\theta}}{c^4(1-s_{13}^2)}.
\end{eqnarray}
From Eq. (\ref{Eq:Jarlskog}), $\delta$ is written as follows:
\begin{widetext}
\begin{eqnarray} 
&&\tan{\delta}=\frac{b(2c^2-b^2+b^2\cos{2\theta})\sin{\phi}}{-b\{b^2+(2c^2-b^2)\cos{2\theta}\}\cos{\phi}+c(a-b)(a+b)\sin{2\theta}},\nonumber\\
&&(a=m^2-n^2,b=2mn,c=m^2+n^2),
\label{Eq:PT1C2_tandelta}
\end{eqnarray}
and 
\begin{eqnarray} 
&&\tan{\delta}=\frac{b(c^2+a^2+b^2\cos{2\theta})\sin{\phi}}{b\{b^2+(c^2+a^2)\cos{2\theta}\}\cos{\phi}+c(b-a)(b+a)\sin{2\theta}},\nonumber\\
&&(a=2mn,b=m^2-n^2,c=m^2+n^2).
\label{Eq:PT2C2_tandelta}
\end{eqnarray}
\end{widetext}

Eqs. (\ref{Eq:PT1C1_tandelta}) and (\ref{Eq:PT2C2_tandelta}), as well as Eqs. (\ref{Eq:PT2C1_tandelta}) and (\ref{Eq:PT1C2_tandelta}), are identical. We will algebraically demonstrate this topic in a future study.
%=================================================
%=================================================
\section{Numerical calculations\label{section:NC}}
%%----------------------------------------------------------------------------------
First, we performed numerical calculations for $m$ and $n$, $\theta$ and $\phi$ within the ranges of $1 \sim1000$, $0$ to $\frac{\pi}{2}$, and $0$ to $2\pi$, searching for combinations that satisfy the observed values. Subsequently, we selected three combinations of $m$ and $n$ and conducted numerical calculations to investigate the dependence of neutrino parameters on $\theta$ and $\phi$. We investigated the relationship between $m$ and $n$ using the Pearson product-moment correlation coefficient (PPMCC)\cite{Pearson1895} as follows:
\begin{eqnarray}
r_{mn} = \frac{{\displaystyle \sum_{k = 1}^l (m_k - \overline{m})
(n_k - \overline{n})}}{\sqrt{{\displaystyle \sum_{k = 1}^l 
(m_k - \overline{m})^2}} \sqrt{{\displaystyle \sum_{k = 1}^l 
(n_k - \overline{n})^2}}}.
\end{eqnarray}
PPMCC measures the linear correlation between two sets of data within the range of
\begin{eqnarray}
-1\leq r_{mn}  \leq 1.
\end{eqnarray}
Positive and negative linear correlations are observed as the value approaches $1$ and $-1$, respectively. If the value is close to $0$, no correlation is observed. $(m_1,n_1),(m_2,n_2),\cdots,(m_k,n_k),\cdots,(m_l,n_l)$ represents $l$ sets of data comprising $m$ and $n$, and it becomes  $(m_1,n_1),(m_2,n_2),\cdots,(m_k,n_k),\cdots,(m_{9716},n_{9716})=(11,6),(15,8),\cdots,(1000,551)$ (Fig. \ref{fig:m_n_PM1C1}). Furthermore, $\overline{m}$ and $\overline{n}$ represent the average, as described in
\begin{eqnarray}
\overline{m}=\frac{{\sum_{k = 1}^l m_k }}{l},
\end{eqnarray}
and
\begin{eqnarray}
\overline{n}=\frac{{\sum_{k = 1}^l n_k }}{l}.
\end{eqnarray}
In the case of Fig. \ref{fig:m_n_PM1C1}, $l$ is $9716$, and the averages of $m$ and $n$ are $\overline{m}=665$ and $\overline{n}=350$.

If PPMC $r_{mn}$ is close to 1 (-1), we employed the linear regression model, 
\begin{eqnarray}
n=w m+\beta,
\label{eq:lineari}
\end{eqnarray}
where $w$ and $\beta$ denote the slope and intercept, respectively.
%--------------------------------------------------------------------
\begin{figure*}[th]
\includegraphics[keepaspectratio, scale=0.7]{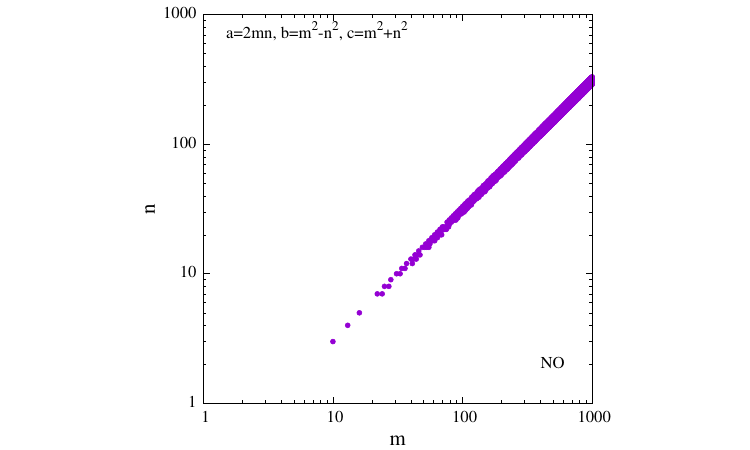}
\includegraphics[keepaspectratio, scale=0.7]{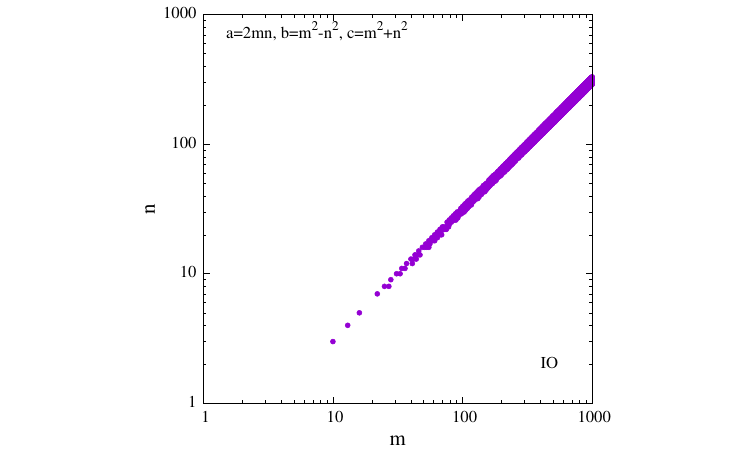}
 \caption{Similar to Fig. \ref{fig:m_n_PM1C1}, except for $\left(a,b,c\right)=\left(2mn, m^2-n^2,  m^2+n^2\right)$ and $\left(U_{\rm{PM}}\right)_{\rm{C1}}$.}
 \label{fig:m_n_PM2C1}
  \end{figure*}
%--------------------------------------------------------------------
%--------------------------------------------------------------------
\begin{figure*}[th]
\includegraphics[keepaspectratio, scale=0.7]{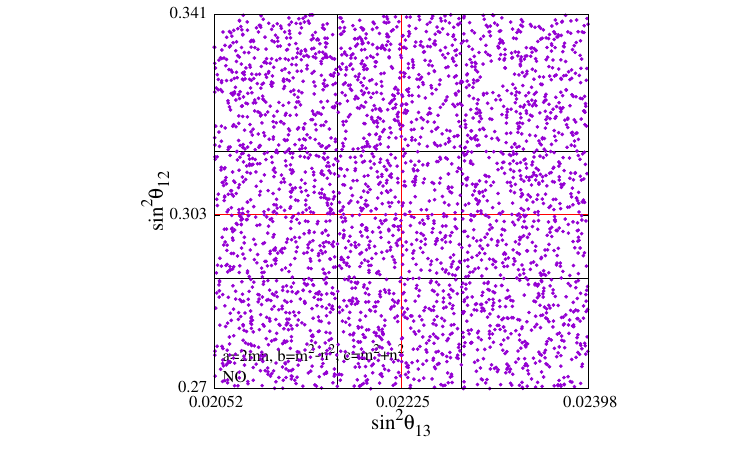}
\includegraphics[keepaspectratio, scale=0.7]{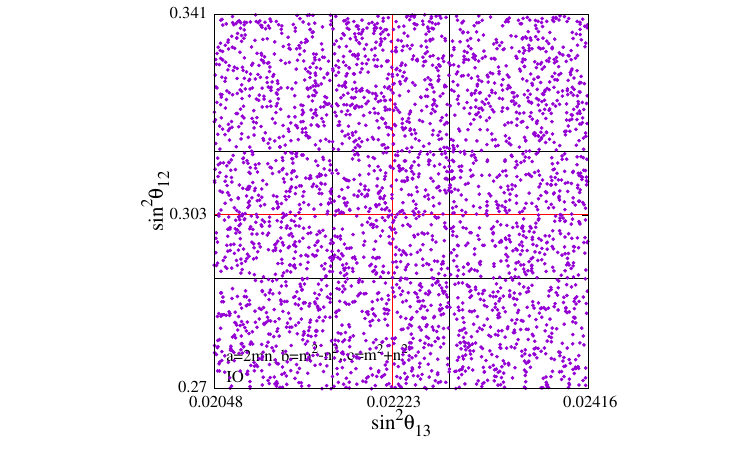}
\includegraphics[keepaspectratio, scale=0.7]{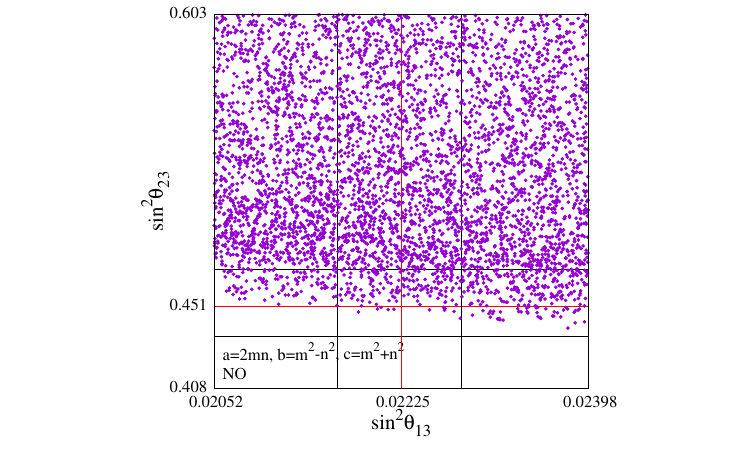}
\includegraphics[keepaspectratio, scale=0.7]{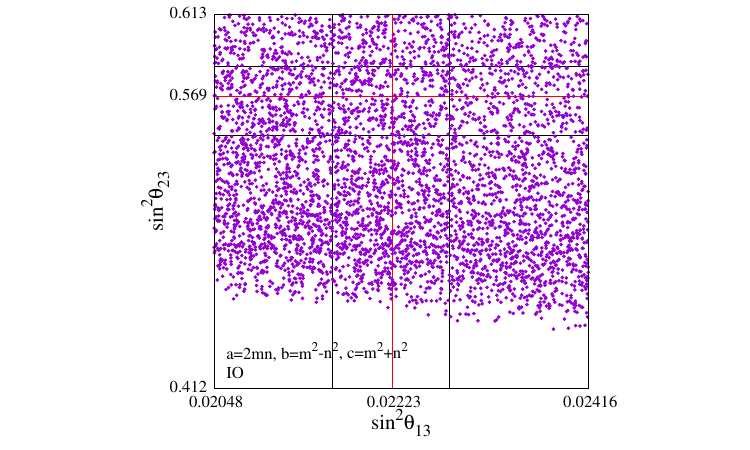}
 \caption{Similar to Fig. \ref{fig:12_13_23_PM1C1}, except for $\left(a,b,c\right)=\left(2mn, m^2-n^2,  m^2+n^2\right)$ and $\left(U_{\rm{PM}}\right)_{\rm{C1}}$.}
 \label{fig:12_13_23_PM2C1}
  \end{figure*}
%--------------------------------------------------------------------
%--------------------------------------------------------------------
\begin{figure*}[t]
\includegraphics[keepaspectratio, scale=0.7]{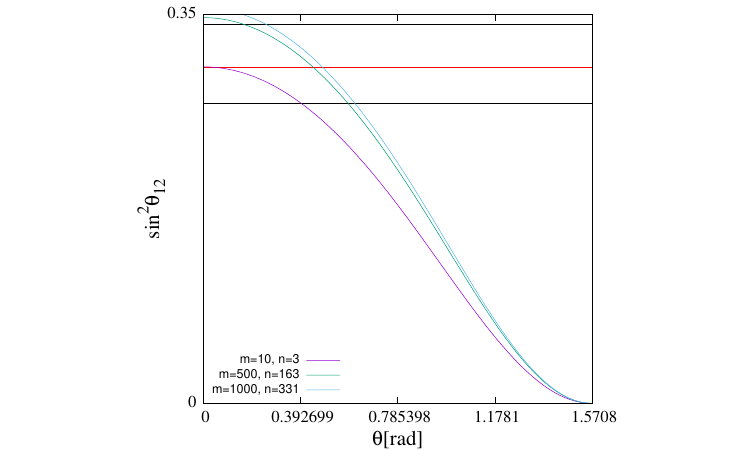}\\
\includegraphics[keepaspectratio, scale=0.7]{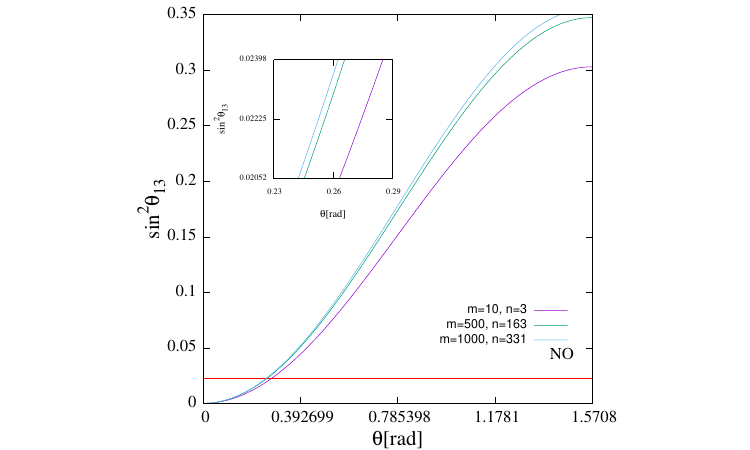}
\includegraphics[keepaspectratio, scale=0.7]{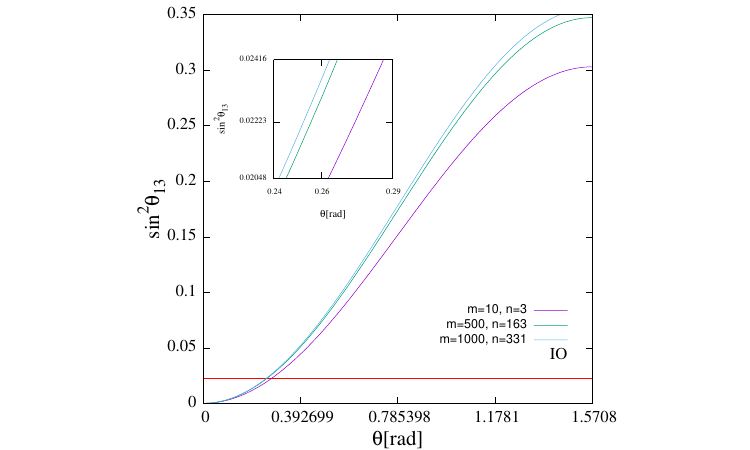}
 \caption{Similar to Fig. \ref{fig:12_13_t_PM1C1}, except for $\left(a,b,c\right)=\left(2mn, m^2-n^2,  m^2+n^2\right)$ and $\left(U_{\rm{PM}}\right)_{\rm{C1}}$. We selected $(m,n)=(10,3),(500,163),(1000,331)$.}
 \label{fig:12_13_t_PM2C1}
  \end{figure*}
%--------------------------------------------------------------------
%--------------------------------------------------------------------
\begin{figure*}[th]
\includegraphics[keepaspectratio, scale=0.7]{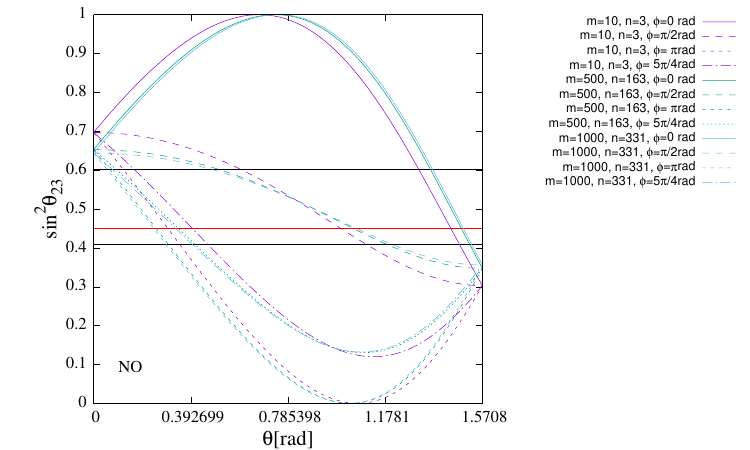}
\includegraphics[keepaspectratio, scale=0.7]{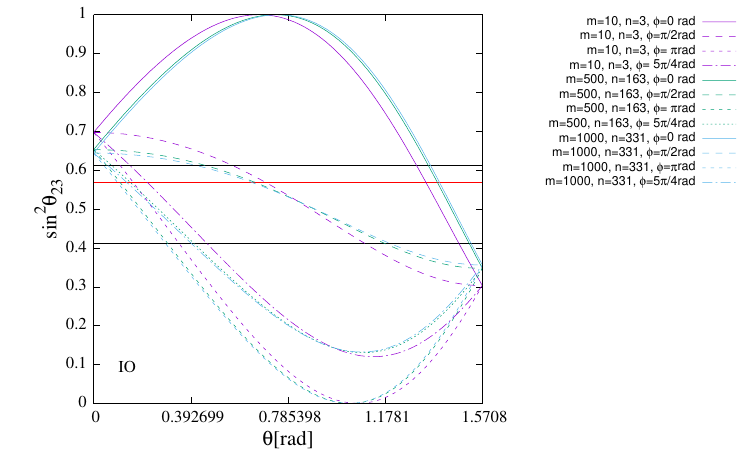}
 \caption{Similar to Fig. \ref{fig:23_t_PM1C1}, except for $\left(a,b,c\right)=\left(2mn, m^2-n^2,  m^2+n^2\right)$ and $\left(U_{\rm{PM}}\right)_{\rm{C1}}$. We selected $(m,n)=(10,3),(500,163),(1000,331)$.}
 \label{fig:23_t_PM2C1}
  \end{figure*}
%--------------------------------------------------------------------
%--------------------------------------------------------------------
\begin{figure*}[th]
\includegraphics[keepaspectratio, scale=0.7]{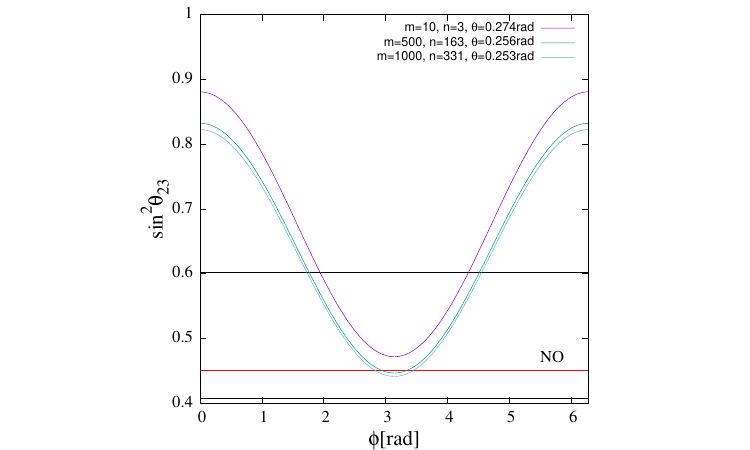}
\includegraphics[keepaspectratio, scale=0.7]{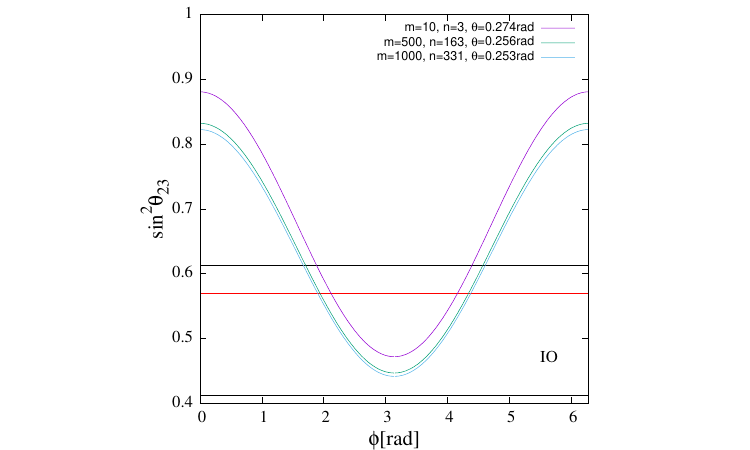}
 \caption{Similar to Fig. \ref{fig:23_p_PM1C1}, except for $\left(a,b,c\right)=\left(2mn, m^2-n^2,  m^2+n^2\right)$ and $\left(U_{\rm{PM}}\right)_{\rm{C1}}$. We selected $(m,n)=(10,3),(500,163),(1000,331)$.}
 \label{fig:23_p_PM2C1}
  \end{figure*}
%--------------------------------------------------------------------
%--------------------------------------------------------------------
\begin{figure*}[th]
\includegraphics[keepaspectratio, scale=0.7]{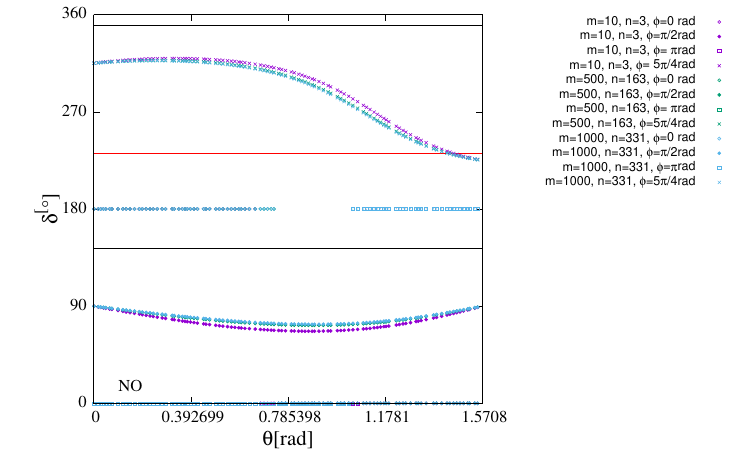}
\includegraphics[keepaspectratio, scale=0.7]{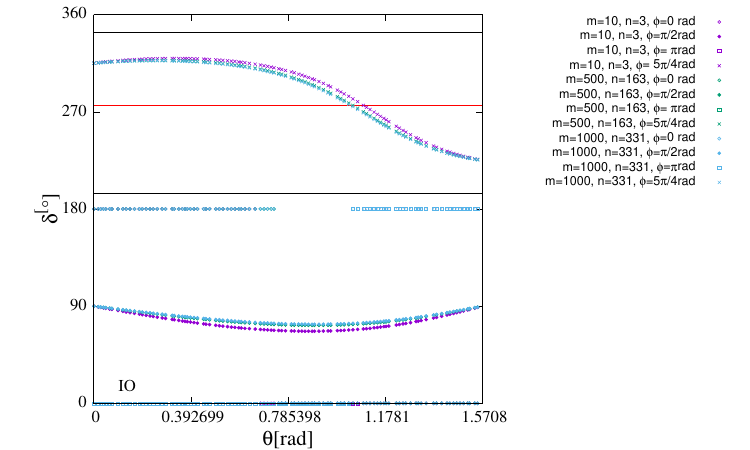}
 \caption{Similar to Fig. \ref{fig:d_t_PM1C1}, except for $\left(a,b,c\right)=\left(2mn, m^2-n^2,  m^2+n^2\right)$ and $\left(U_{\rm{PM}}\right)_{\rm{C1}}$. We selected $(m,n)=(10,3),(500,163),(1000,331)$.}
 \label{fig:d_t_PM2C1}
  \end{figure*}
%--------------------------------------------------------------------
%--------------------------------------------------------------------
\begin{figure*}[th]
\includegraphics[keepaspectratio, scale=0.7]{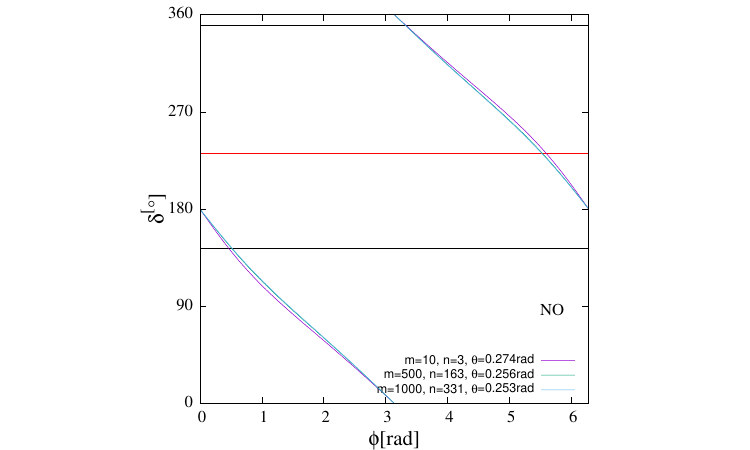}
\includegraphics[keepaspectratio, scale=0.7]{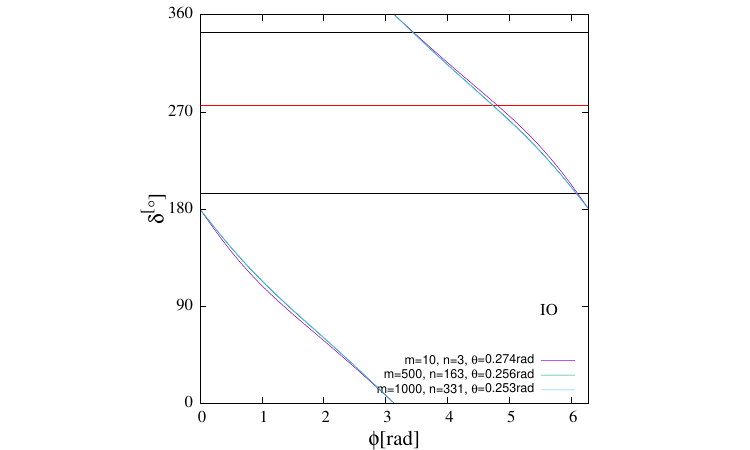}
 \caption{Similar to Fig. \ref{fig:d_p_PM1C1}, except for $\left(a,b,c\right)=\left(2mn, m^2-n^2,  m^2+n^2\right)$ and $\left(U_{\rm{PM}}\right)_{\rm{C1}}$. We selected $(m,n)=(10,3),(500,163),(1000,331)$.}
 \label{fig:d_p_PM2C1}
  \end{figure*}
%--------------------------------------------------------------------
%%----------------------------------------------------------------------------------
\subsection{$\left(U_{\rm{PM}}\right)_{\rm{C1}}$\label{section:PMC1}}
%%----------------------------------------------------------------------------------
%%----------------------------------------------------------------------------------
\subsubsection{The case of the $\left(a,b,c\right)=\left(m^2-n^2, 2mn, m^2+n^2\right)$\label{section:PM1C1}}
%%----------------------------------------------------------------------------------
Fig. \ref{fig:m_n_PM1C1} illustrates the dependence of the $m$ on $n$ for $\left(U_{\rm{PM}}\right)_{\rm{C1}}$ with the use of $\left(a,b,c\right)=$

$\left(m^2-n^2, 2mn, m^2+n^2\right)$. We varied $\theta$ and $\phi$ within the ranges of $0$ to $\frac{\pi}{2}$, and $0$ to $2\pi$, respectively.

Fig. \ref{fig:m_n_PM1C1} shows $m$ and $n$ satisfying the $3 \sigma$ region for mixing angles, $s_{ij}^2$, and the Dirac CP phase, $\delta$. The left and right panels shows the NO and IO cases, respectively. The key observations from Fig. \ref{fig:m_n_PM1C1} are as follows:
\begin{itemize}
\item In the NO and IO cases, the PPMCs were both $0.997$. Considering that both PPMCs were close to $1$, a very strong positive linear correlation was observed between $m$ and $n$.
\item The relationship between $m$ and $n$ exhibits strong linearity. Thus, using Eq. (\ref{eq:lineari}), the slope, $w$, was $0.527$ and $0.526$ in the NO and IO cases, respectively. If $m$ increased by $10$, $n$ increased by $5$. The intercept, $\beta$, was $-0.0869$ and $-0.104$ in the NO and IO cases, respectively. Slope $w$ and intercept $\beta$ did not significantly vary between the NO and IO cases. 
\end{itemize}

Fig. \ref{fig:12_13_23_PM1C1} illustrates the relationship between $s_{12}^2$ and $s_{13}^2$, in addition to the relationship between $s_{23}^2$ and $s_{13}^2$, for $\left(U_{\rm{PM}}\right)_{\rm{C1}}$ using $\left(a, b, c\right)=\left(m^2-n^2, 2mn, m^2+n^2\right)$, satisfying the $3 \sigma$ regions of mixing angles $s_{ij}^2$ and the Dirac CP phase, $\delta$.  We varied $\theta$,  $\phi$, and $m$ and $n$ within the ranges of $0$ to $\frac{\pi}{2}$, $0$ to $2\pi$, and $0 \sim1000$, respectively.

The upper (lower) panel shows the relationship between $s_{12}^2$ and $s_{13}^2$ ($s_{23}^2$ and $s_{13}^2$). The upper and lower left panels show the NO case, and the upper and lower right panels show the IO case. The red line in the figure represents the best-fit values of $s_{12}^2$, $s_{13}^2$, and $s_{23}^2$, and the black line shows the $3 \sigma$ region for $s_{12}^2$, $s_{13}^2$, and $s_{23}^2$.
\begin{itemize}
 \item In the NO and IO cases, the predicted values of $s_{12}^2$, $s_{13}^2$, and $s_{23}^2$ from the model fell within the $3 \sigma$ region.
 \item If we appropriately select $m$, $n$, and $\theta$, the predicted values of $s_{12}^2$,$s_{13}^2$, and $s_{23}^2$ from the model can simultaneously satisfy the best-fit values.
\end{itemize}

A benchmark point,
\begin{eqnarray}
(m, n, \theta, \phi) = (11,6,0.2892~\rm{rad}, 3.364~ \rm{rad})
\label{Eq:PM1C1_p}
\end{eqnarray}
yields
\begin{eqnarray}
(s_{12}^2,s_{13}^2,s_{23}^2,\delta) = (0.2759, 0.02384, 0.4735, 191.6^\circ).
\label{Eq:PM1C1_mixing_angle}
\end{eqnarray}

Next, using $(m,n)=(11,6),(502,265),(1000,551)$, we investigated the relationship between neutrino mixing angles, $\delta$, rotation angle $\theta$, and phase parameter $\phi$.

Fig. \ref{fig:12_13_t_PM1C1} illustrates the dependence of the $s_{12}^2$ and $s_{13}^2$ on $\theta$ in the cases of $m=11,n=6$, $m=502, n=265$ and $m=1000,n=551$. The red line represents the best-fit values of $s_{12}^2$ and $s_{13}^2$, and the black line in the upper panel indicates the $3 \sigma$ region for $s_{12}^2$. The enlarged view in the lower panel illustrates the dependence of $s_{13}^2$ on $\theta$ when expanded into the 3$\sigma$ region of $s_{
13}^2$. The key observations from Fig. \ref{fig:12_13_t_PM1C1} are as follows:
\begin{itemize}
\item $\theta$ exists to satisfy  the 3$\sigma$ region of $s_{13}^2$ and $s_{12}^2$.
 \item The range of $\theta$ satisfying the 3$\sigma$ region of $s_{13}^2$ was limited to $0.255~\rm{rad}\leq \theta \leq 0.296~\rm{rad}$.
  \item If the range of $\theta$ was $0.255~\rm{rad}\leq \theta \leq 0.296~\rm{rad}$, $s_{12}^2$ was within the 3$\sigma$ region of $s_{12}^2$.
\end{itemize}

Fig. \ref{fig:23_t_PM1C1} illustrates the dependence of $s_{23}^2$ on $\theta$ in the cases of $m=11,n=6$, $m=502, n=265$ and $m=1000,n=551$. We selected $0$,$\frac{\pi}{2}$,$\pi$,$\frac{5\pi}{4}$ for $\phi$. The red line represents the best-fit values of $s_{23}^2$, and the black line shows the $3\sigma$ region for $s_{23}^2$. The left and right panels show the NO and IO cases, respectively. 

Fig. \ref{fig:23_p_PM1C1} illustrates the dependence of $s_{23}^2$ on $\phi$ in the cases of $m=11,n=6$, $m=502, n=265$ and $m=1000,n=551$. We selected $\theta$ such that $s_{13}^2$ was approximately $0.022$. The left and right panels show the NO and IO cases, respectively. The key observations from Figs.\ref{fig:23_t_PM1C1} and \ref{fig:23_p_PM1C1} are as follows:
\begin{itemize}
 \item $\theta$ and $\phi$ exist to satisfy  the 3$\sigma$ region of $s_{23}^2$.
 \item This combination of $m$ and $n$ does not satisfy the best-fit value for NO; however, it satisfies the best-fit value for IO.
 \item The other $m$ and $n$ combinations satisfy the best-fit value of NO. For example, a benchmark point, $(m, n, \theta, \phi) = (44,23,0.2743~\rm{rad}, 3.188~ \rm{rad})$, yields $(s_{12}^2,s_{13}^2,s_{23}^2,\delta)=$ 
 
 $(0.3093, 0.02390, 0.4512, 182.5^\circ)$.
  \item As shown in Fig. \ref{fig:23_p_PM1C1}, the range of $\phi$ satisfying the 3$\sigma$ region of $s_{23}^2$ was limited to $1.8~\rm{rad} \leq \phi \leq 4.5~\rm{rad}$.
\end{itemize}

Fig. \ref{fig:d_t_PM1C1} illustrates the dependence of $\delta$ on $\theta$ in the cases of $m=11,n=6$, $m=502, n=265$ and $m=1000,n=551$. We selected $0$,$\frac{\pi}{2}$,$\pi$,$\frac{5\pi}{4}$ for $\phi$. The red line represents the best-fit values of $\delta$, and the black line shows the $3 \sigma$ region for $\delta$. The left and right panels show the NO and IO cases, respectively.

Fig. \ref{fig:d_p_PM1C1} illustrates the dependence of $\delta$ on $\phi$ in the cases of $m=11,n=6$, $m=502, n=265$ and $m=1000,n=551$. We selected $\theta$ such that $s_{13}^2$ was approximately $0.022$. The left and right panels show the NO and IO cases, respectively. The key observations from Figs.\ref{fig:d_t_PM1C1} and \ref{fig:d_p_PM1C1} are as follows:
\begin{itemize}
 \item No difference was observed in the $\delta$ for the different combinations of $m$ and $n$.
 \item As shown in Fig. \ref{fig:d_p_PM1C1}, the range of $\phi$ satisfying the 3$\sigma$ region of $\delta$ was limited to $2.4~\rm{rad} \leq\phi\leq6.2~\rm{rad}$.
\end{itemize}
%%----------------------------------------------------------------------------------
\subsubsection{The case of the $\left(a,b,c\right)=\left(2mn, m^2-n^2,  m^2+n^2\right)$\label{section:PM2C1}}
%%----------------------------------------------------------------------------------
Fig. \ref{fig:m_n_PM2C1} is the same as  Fig. \ref{fig:m_n_PM1C1}, except for $\left(a,b,c\right)=\left(2mn, m^2-n^2,  m^2+n^2\right)$ and $\left(U_{\rm{PM}}\right)_{\rm{C1}}$. The key observations from Fig. \ref{fig:m_n_PM2C1} are as follows:
\begin{itemize}
 \item In the NO and IO cases, the PPMCs were both $0.994$. The PPMCs
were close to $1$; thus, a very strong positive linear correlation was observed between $m$ and $n$.
 \item The relationship between $m$ and $n$ exhibits strong linearity. Thus, using Eq. (\ref{eq:lineari}), slope $w$ was $0.311$ in the NO and IO cases, respectively. If $m$ increased by $10$, $n$ increased by $3$. The intercept $\beta$ was $-0.150$ and $-0.196$ in the NO and IO cases, respectively. Slope $w$ and intercept $\beta$ did not significantly vary between the NO and IO cases.
 \item The slope of Fig. \ref{fig:m_n_PM2C1} is smaller than that of Fig. \ref{fig:m_n_PM1C1}.
\end{itemize}

Fig. \ref{fig:12_13_23_PM2C1} is the same as Fig. \ref{fig:12_13_23_PM1C1}, except, for $\left(a,b,c\right)=\left(2mn, m^2-n^2,  m^2+n^2\right)$ and $\left(U_{\rm{PM}}\right)_{\rm{C1}}$. The key observations from Fig. \ref{fig:12_13_23_PM2C1} are as follows:
\begin{itemize}
 \item In the NO and IO cases, the predicted values of $s_{12}^2$, $s_{13}^2$, and $s_{23}^2$ from the model fell within the $3 \sigma$ region.
 \item If we appropriately selected $m$, $n$ and $\theta$, the predicted values of $s_{12}^2$, $s_{13}^2$, and $s_{23}^2$ from the model can simultaneously satisfy the best-fit values.
\end{itemize}

A benchmark point,
\begin{eqnarray}
(m, n, \theta, \phi) = (10,3,0.2785~\rm{rad}, 3.952~ \rm{rad}),
\label{Eq:PM1C1_p}
\end{eqnarray}
yields
\begin{eqnarray}
(s_{12}^2,s_{13}^2,s_{23}^2,\delta) = (0.2867, 0.02290, 0.5327, 318.1^\circ).
\label{Eq:PM1C1_mixing_angle}
\end{eqnarray}

Next, using $(m,n)=(10,3),(500,163),(1000,331)$, we investigated the relationship between the neutrino mixing angles, $\delta$, rotation angle $\theta$ and phase parameter $\phi$.

Fig. \ref{fig:12_13_t_PM2C1} is the same as  Fig. \ref{fig:12_13_t_PM1C1}, except for $\left(a,b,c\right)=\left(2mn, m^2-n^2,  m^2+n^2\right)$ and $\left(U_{\rm{PM}}\right)_{\rm{C1}}$. The key observations from Fig. \ref{fig:12_13_t_PM2C1} are as follows:

\begin{itemize}
\item $\theta$ exists to satisfy the 3$\sigma$ region of $s_{13}^2$ and $s_{12}^2$.
 \item The range of $\theta$ satisfying the 3$\sigma$ region of $s_{13}^2$ was limited to $0.241~\rm{rad}\leq\theta\leq0.285~\rm{rad}$.
 \item If the range of $\theta$ is $0.241~\rm{rad}\leq\theta\leq0.285~\rm{rad}$, $s_{12}^2$ is within the 3$\sigma$ region of $s_{12}^2$.
\end{itemize}

Fig. \ref{fig:23_t_PM2C1} is the same as  Fig. \ref{fig:23_t_PM1C1}, except for $\left(a,b,c\right)=\left(2mn, m^2-n^2,  m^2+n^2\right)$ and $\left(U_{\rm{PM}}\right)_{\rm{C1}}$. Fig. \ref{fig:23_p_PM2C1} is the same as  Fig. \ref{fig:23_p_PM1C1}, except for $\left(a,b,c\right)=\left(2mn, m^2-n^2,  m^2+n^2\right)$ and $\left(U_{\rm{PM}}\right)_{\rm{C1}}$. The key observations from Figs. \ref{fig:23_t_PM2C1} and \ref{fig:23_p_PM2C1} are as follows:
\begin{itemize}
 \item $\theta$ and $\phi$ exist to satisfy the 3$\sigma$ region of $s_{23}^2$.
 \item There is a combination of $m$ and $n$ that satisfies the best-fit values of NO and IO.
  \item As shown in Fig. \ref{fig:23_p_PM2C1}, the range of $\phi$ satisfying the  3$\sigma$ region of $s_{23}^2$ was limited to $1.7~\rm{rad} \leq\phi\leq4.6~\rm{rad}$.
\end{itemize}

Fig. \ref{fig:d_t_PM2C1} is the same as  Fig. \ref{fig:d_t_PM1C1}, except for $\left(a,b,c\right)=\left(2mn, m^2-n^2,  m^2+n^2\right)$ and $\left(U_{\rm{PM}}\right)_{\rm{C1}}$. Fig. \ref{fig:d_p_PM2C1} is the same as  Fig. \ref{fig:d_p_PM1C1}, except for $\left(a,b,c\right)=\left(2mn, m^2-n^2,  m^2+n^2\right)$ and $\left(U_{\rm{PM}}\right)_{\rm{C1}}$. The key observations from Fig. \ref{fig:d_t_PM2C1} and \ref{fig:d_p_PM2C1} are as follows: 
\begin{itemize}
 \item No difference was observed in the $\delta$ for different combinations of $m$ and $n$.
 \item As shown in Fig. \ref{fig:d_p_PM2C1}, the range of $\phi$ satisfying the 3$\sigma$ region of $\delta$ was limited to $0~\rm{rad} \leq\phi\leq0.5~\rm{rad}$ and $3.3~\rm{rad} \leq\phi\leq2\pi~\rm{rad}$.
\end{itemize}
%%----------------------------------------------------------------------------------
\subsection{$\left(U_{\rm{PM}}\right)_{\rm{C2}}$\label{section:PMC2}}
%%----------------------------------------------------------------------------------
%%----------------------------------------------------------------------------------
\subsubsection{The case of the $\left(a,b,c\right)=\left(m^2-n^2, 2mn, m^2+n^2\right)$\label{section:PM1C2}}
%%----------------------------------------------------------------------------------
%--------------------------------------------------------------------
\begin{figure*}[t]
\includegraphics[keepaspectratio, scale=0.7]{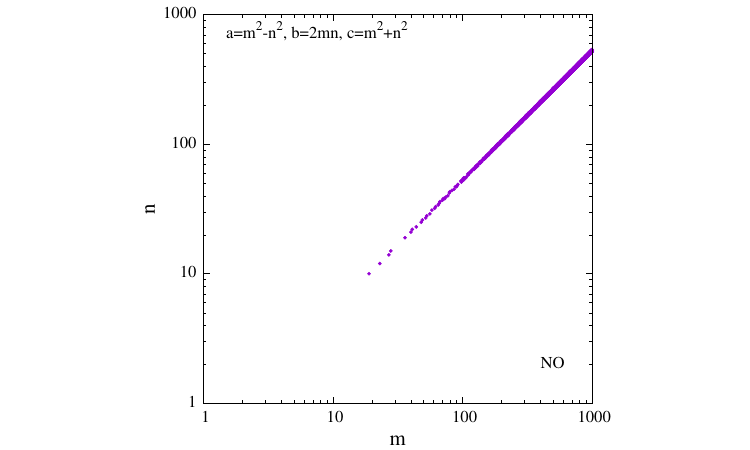}
\includegraphics[keepaspectratio, scale=0.7]{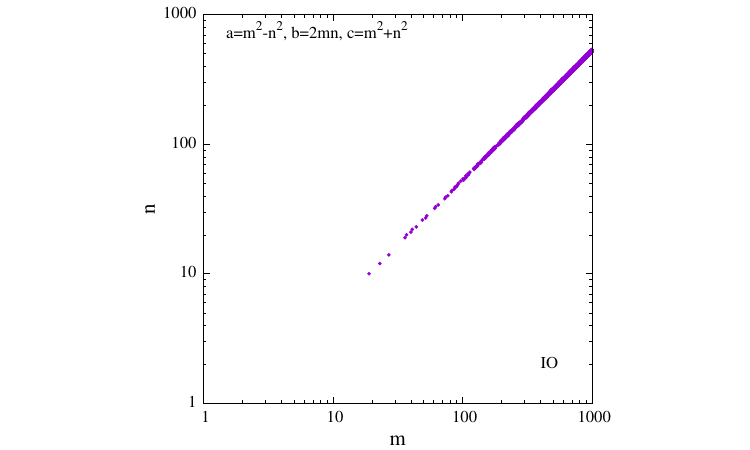}
 \caption{Similar to Fig. \ref{fig:m_n_PM1C1}, except for $\left(a,b,c\right)=\left(m^2-n^2, 2mn, m^2+n^2\right)$ and $\left(U_{\rm{PM}}\right)_{\rm{C2}}$.}
 \label{fig:m_n_PM1C2}
  \end{figure*}
%--------------------------------------------------------------------
%--------------------------------------------------------------------
\begin{figure*}[th]
\includegraphics[keepaspectratio, scale=0.7]{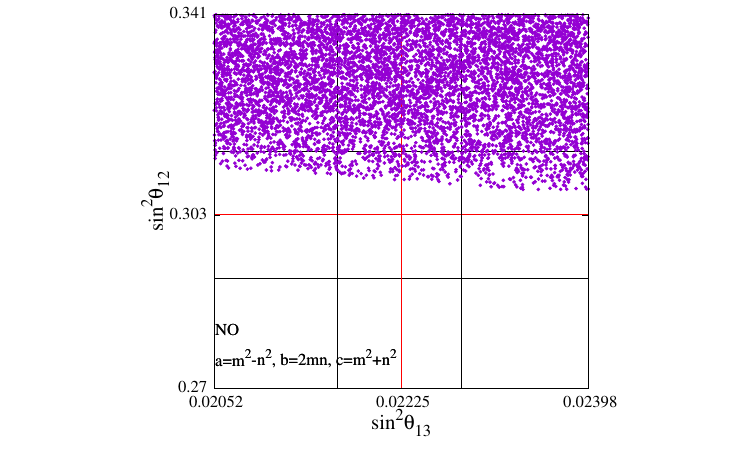}
\includegraphics[keepaspectratio, scale=0.7]{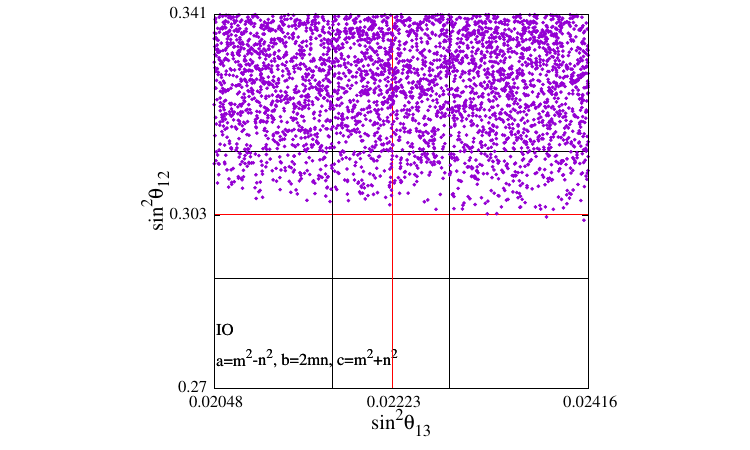}
\includegraphics[keepaspectratio, scale=0.7]{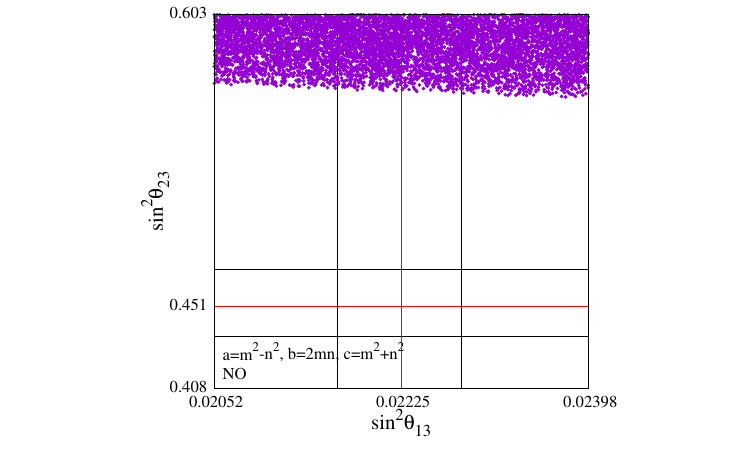}
\includegraphics[keepaspectratio, scale=0.7]{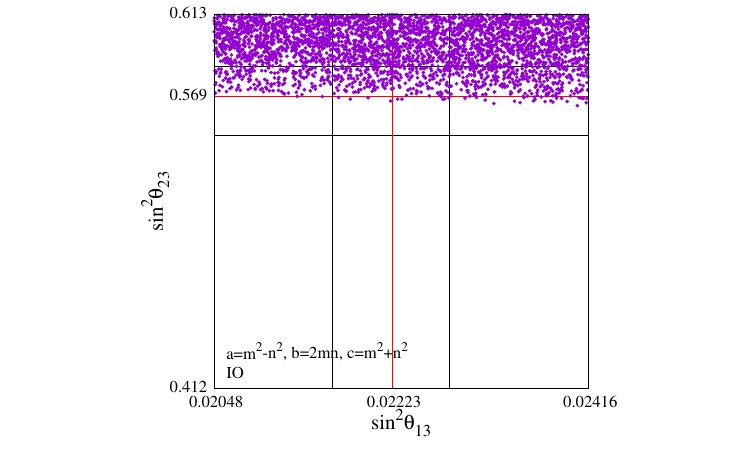}
 \caption{Similar to Fig. \ref{fig:12_13_23_PM1C1}, except for $\left(a,b,c\right)=\left(m^2-n^2, 2mn, m^2+n^2\right)$ and $\left(U_{\rm{PM}}\right)_{\rm{C2}}$.}
 \label{fig:12_13_23_PM1C2}
  \end{figure*}
%--------------------------------------------------------------------
%--------------------------------------------------------------------
\begin{figure*}[t]
\includegraphics[keepaspectratio, scale=0.7]{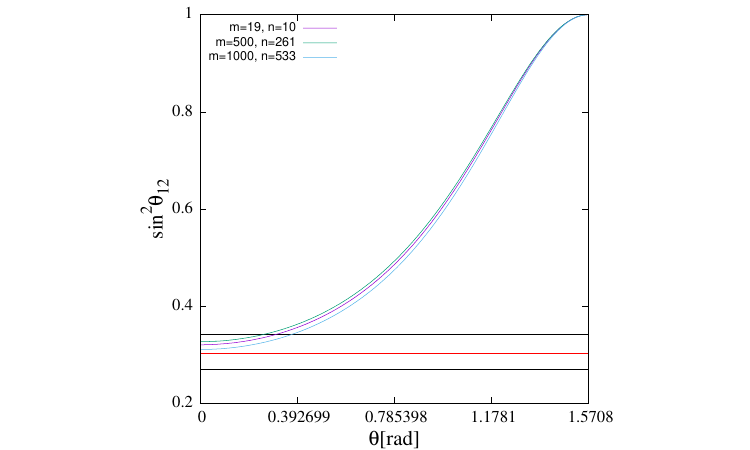}\\
\includegraphics[keepaspectratio, scale=0.7]{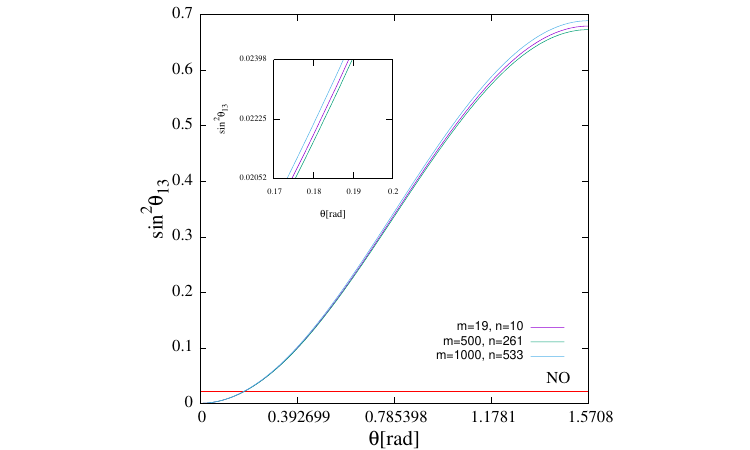}
\includegraphics[keepaspectratio, scale=0.7]{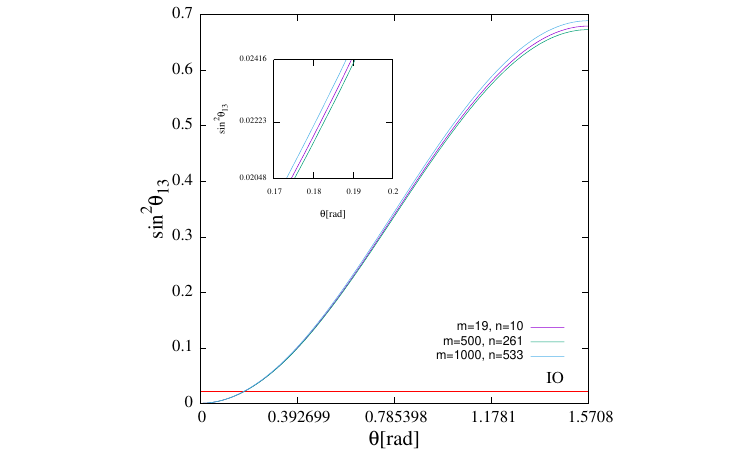}
 \caption{Similar to Fig. \ref{fig:12_13_t_PM1C1}, except for $\left(a,b,c\right)=\left(m^2-n^2, 2mn, m^2+n^2\right)$ and $\left(U_{\rm{PM}}\right)_{\rm{C2}}$. We selected $(m,n)=(19,10),(500,261),(1000,553)$.}
 \label{fig:12_13_t_PM1C2}
  \end{figure*}
%--------------------------------------------------------------------
%--------------------------------------------------------------------
\begin{figure*}[th]
\includegraphics[keepaspectratio, scale=0.7]{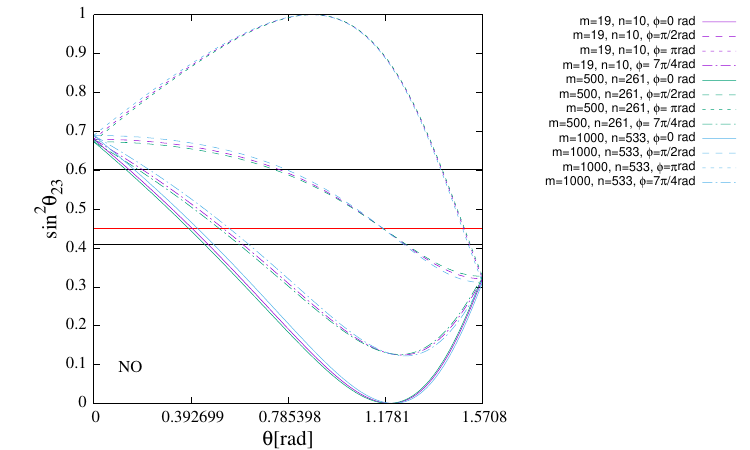}
\includegraphics[keepaspectratio, scale=0.7]{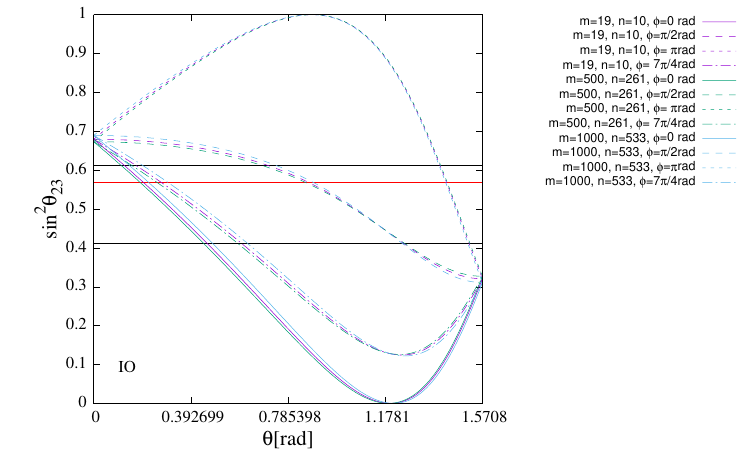}
 \caption{Similar to Fig. \ref{fig:23_t_PM1C1}, except for $\left(a,b,c\right)=\left(m^2-n^2, 2mn, m^2+n^2\right)$ and $\left(U_{\rm{PM}}\right)_{\rm{C2}}$. We selected $(m,n)=(19,10),(500,261),(1000,553)$ and $\phi=0, \frac{\pi}{2},\pi,\frac{7\pi}{4}$.}
 \label{fig:23_t_PM1C2}
  \end{figure*}
%--------------------------------------------------------------------
%--------------------------------------------------------------------
\begin{figure*}[th]
\includegraphics[keepaspectratio, scale=0.7]{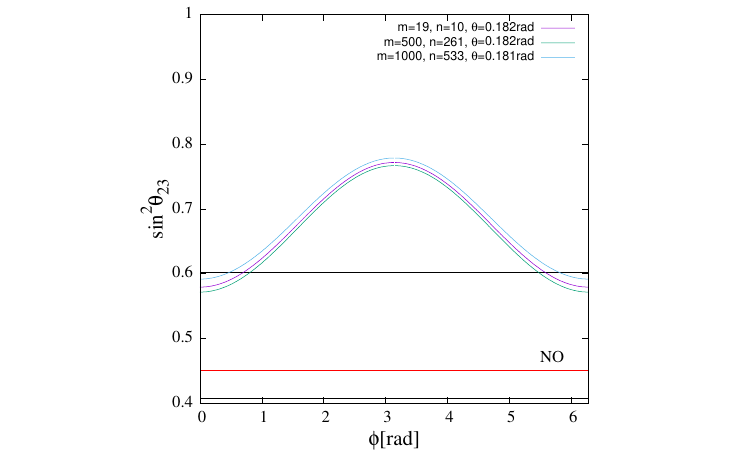}
\includegraphics[keepaspectratio, scale=0.7]{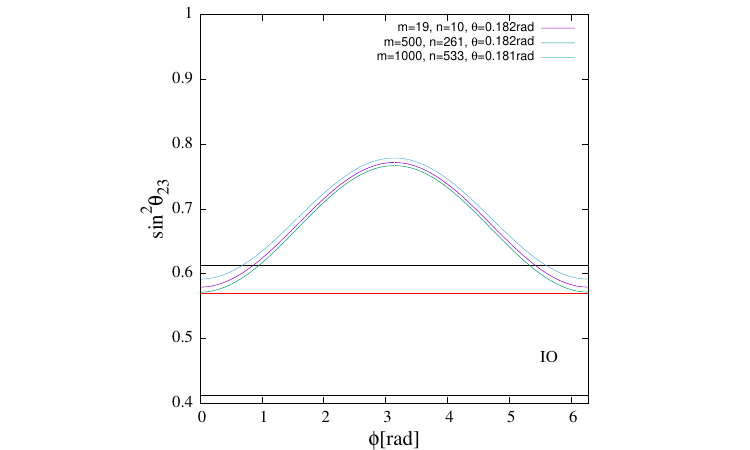}
 \caption{Similar to Fig. \ref{fig:23_p_PM1C1}, except for $\left(a,b,c\right)=\left(m^2-n^2, 2mn, m^2+n^2\right)$ and $\left(U_{\rm{PM}}\right)_{\rm{C2}}$.We selected $(m,n)=(19,10),(500,261),(1000,553)$.}
 \label{fig:23_p_PM1C2}
  \end{figure*}
%--------------------------------------------------------------------
%--------------------------------------------------------------------
\begin{figure*}[th]
\includegraphics[keepaspectratio, scale=0.7]{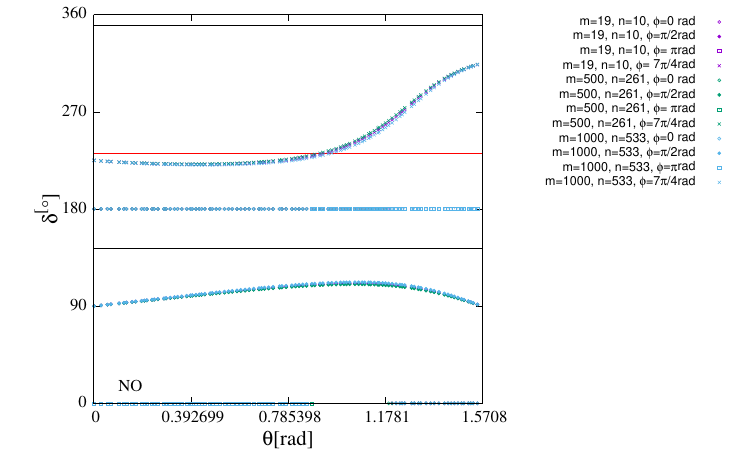}
\includegraphics[keepaspectratio, scale=0.7]{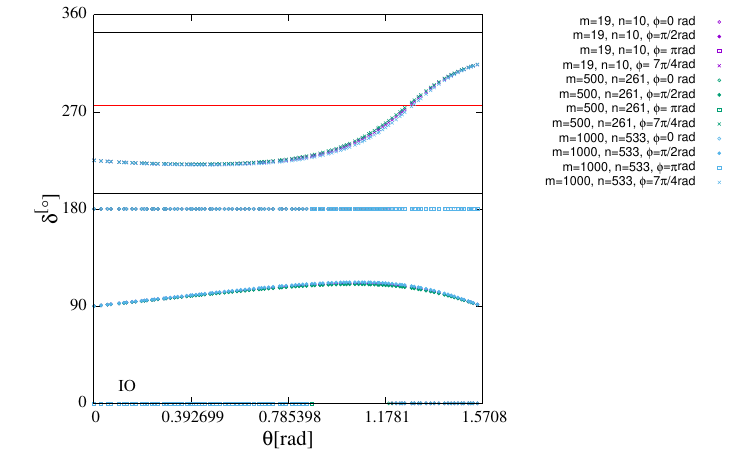}
 \caption{Similar to Fig. \ref{fig:d_t_PM1C1}, except for $\left(a,b,c\right)=\left(m^2-n^2, 2mn, m^2+n^2\right)$ and $\left(U_{\rm{PM}}\right)_{\rm{C2}}$. We selected $(m,n)=(19,10),(500,261),(1000,553)$ and $\phi=0, \frac{\pi}{2},\pi,\frac{7\pi}{4}$.}
 \label{fig:d_t_PM1C2}
  \end{figure*}
%--------------------------------------------------------------------
%--------------------------------------------------------------------
\begin{figure*}[th]
\includegraphics[keepaspectratio, scale=0.7]{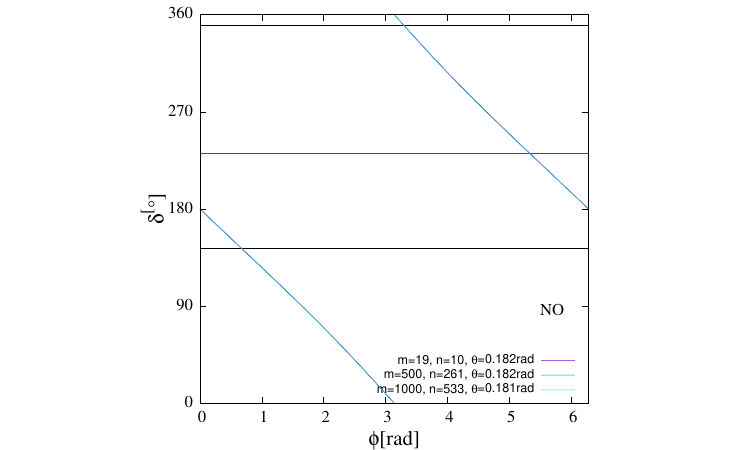}
\includegraphics[keepaspectratio, scale=0.7]{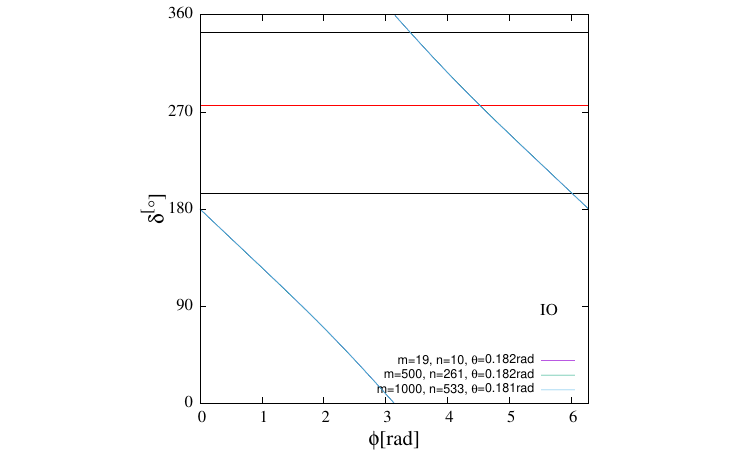}
 \caption{Similar to Fig. \ref{fig:d_p_PM1C1}, except for $\left(a,b,c\right)=\left(m^2-n^2, 2mn, m^2+n^2\right)$ and $\left(U_{\rm{PM}}\right)_{\rm{C2}}$. We selected $(m,n)=(19,10),(500,261),(1000,553)$.}
 \label{fig:d_p_PM1C2}
  \end{figure*}
%--------------------------------------------------------------------
Fig. \ref{fig:m_n_PM1C2} is the same as  Fig. \ref{fig:m_n_PM1C1}, except for $\left(a,b,c\right)=\left(m^2-n^2, 2mn, m^2+n^2\right)$ and $\left(U_{\rm{PM}}\right)_{\rm{C2}}$. The key observations from Fig. \ref{fig:m_n_PM1C2} are as follows:
\begin{itemize}
  \item In the NO and IO cases, the PPMCs were both $0.999$. Both PPMCs were close to $1$; thus, a very strong positive linear correlation was observed between $m$ and $n$.
 \item The relationship between $m$ and $n$ exhibits strong linearity. Thus, using Eq. (\ref{eq:lineari}), $w$ was $0.527$ in both cases. If $m$ increased by $10$, $n$ increased by $5$. $\beta$ was $-0.106$ and $-0.0698$ in the NO and IO cases. Slope $w$ and intercept $\beta$ did not significantly vary between the NO and IO cases.
\end{itemize}

Fig. \ref{fig:12_13_23_PM1C2} is the same as Fig. \ref{fig:12_13_23_PM1C1}, except for $\left(a,b,c\right)=\left(m^2-n^2, 2mn, m^2+n^2\right)$ and $\left(U_{\rm{PM}}\right)_{\rm{C2}}$. The key observations from Fig. \ref{fig:12_13_23_PM1C2} are as follows:
\begin{itemize}
 \item In the NO and IO cases, the predicted values of $s_{12}^2$, $s_{13}^2$, and $s_{23}^2$ from the model fell within the $3 \sigma$ region.
 \item The predicted values from the model exceeded or were equal to $0.3$ and $0.55$ for $s_{12}^2$ and $s_{23}^2$, respectively, in both cases. Thus, it was impossible to simultaneously satisfy the best-fit values for $s_{12}^2$, $s_{13}^2$ and $s_{23}^2$ in the case of NO.
\end{itemize}

A benchmark point,
\begin{eqnarray}
(m, n, \theta, \phi) = (19,10,0.1751~\rm{rad}, 0.4214~ \rm{rad})
\label{Eq:PM1C1_p}
\end{eqnarray}
yields
\begin{eqnarray}
(s_{12}^2,s_{13}^2,s_{23}^2,\delta) = (0.3273, 0.02062, 0.5914, 157.1^\circ).
\label{Eq:PM1C1_mixing_angle}
\end{eqnarray}

Next, using $(m,n)=(19,10),(500,261),(1000,553)$, we investigated the relationship
between the neutrino mixing angles, $\delta$, rotation angle $\theta$, and phase parameter $\phi$.

Fig. \ref{fig:12_13_t_PM1C2} is the same as Fig. \ref{fig:12_13_t_PM1C1}, except for $\left(a,b,c\right)=\left(2mn, m^2-n^2,  m^2+n^2\right)$ and $\left(U_{\rm{PM}}\right)_{\rm{C1}}$. The key observations from Fig. \ref{fig:12_13_t_PM1C2} are as follows:
\begin{itemize}
\item $\theta$ exists to satisfy the 3$\sigma$ region of $s_{13}^2$ and $s_{12}^2$.
 \item The range of $\theta$ satisfying the 3$\sigma$ region of $s_{13}^2$ was limited to $0.173~\rm{rad}\leq\theta\leq0.187~\rm{rad}$.
 \item If the range of $\theta$ was $0.173~\rm{rad}\leq\theta\leq0.187~\rm{rad}$, $s_{12}^2$ was within the 3$\sigma$ region of $s_{12}^2$. However, it exceeded the best-fit value.
\end{itemize}

Fig. \ref{fig:23_t_PM1C2} is the same as  Fig. \ref{fig:23_t_PM1C1}, except for $\left(a,b,c\right)=\left(2mn, m^2-n^2,  m^2+n^2\right)$ and $\left(U_{\rm{PM}}\right)_{\rm{C1}}$. Fig. \ref{fig:23_p_PM1C2} is the same as  Fig. \ref{fig:23_p_PM1C1}, except for $\left(a,b,c\right)=\left(2mn, m^2-n^2,  m^2+n^2\right)$ and $\left(U_{\rm{PM}}\right)_{\rm{C1}}$. The key observations from Figs.\ref{fig:23_t_PM1C2} and \ref{fig:23_p_PM1C2} are as follows:
\begin{itemize}
 \item $\theta$ and $\phi$ exist to satisfy the 3$\sigma$ region of $s_{23}^2$. 
 \item This combination of $m$ and $n$ satisfies the best-fit value for IO.
 \item This combination of $m$ and $n$ does not satisfy the best-fit value for NO.
 \item If $s_{13}^2$ aligns with the best-fit value, $s_{23}^2$ was close to the upper limit of the 3$\sigma$ region.
  \item As shown in Fig. \ref{fig:23_p_PM1C2}, the range of $\phi$ satisfying the 3$\sigma$ region of $s_{23}^2$ was limited to $0~\rm{rad} \leq\phi\leq1~\rm{rad}$ and $5.3~\rm{rad} \leq\phi\leq2\pi~\rm{rad}$.
\end{itemize}

Fig. \ref{fig:d_t_PM1C2} is the same as  Fig. \ref{fig:d_t_PM1C1}, except for $\left(a,b,c\right)=\left(2mn, m^2-n^2,  m^2+n^2\right)$ and $\left(U_{\rm{PM}}\right)_{\rm{C2}}$. Fig. \ref{fig:d_p_PM1C2} is the same as  Fig. \ref{fig:d_p_PM1C1}, except for $\left(a,b,c\right)=\left(2mn, m^2-n^2,  m^2+n^2\right)$ and $\left(U_{\rm{PM}}\right)_{\rm{C2}}$. The key observations from Figs.\ref{fig:d_t_PM1C2} and \ref{fig:d_p_PM1C2} are as follows:
\begin{itemize}
 \item No difference was observed in the $\delta$ for different combinations of $m$ and $n$.
 \item As shown in Fig. \ref{fig:d_p_PM1C2}, the range of $\phi$ satisfying the 3$\sigma$ region of the $\delta$ was limited to $0~\rm{rad} \leq\phi\leq0.6~\rm{rad}$ and $3.3~\rm{rad} \leq\phi\leq2\pi~\rm{rad}$.
\end{itemize}

%%----------------------------------------------------------------------------------
\subsubsection{The case of the $\left(a,b,c\right)=\left(2mn, m^2-n^2,  m^2+n^2\right)$\label{section:PM2C2}}
%%----------------------------------------------------------------------------------
%--------------------------------------------------------------------
\begin{figure*}[t]
\includegraphics[keepaspectratio, scale=0.7]{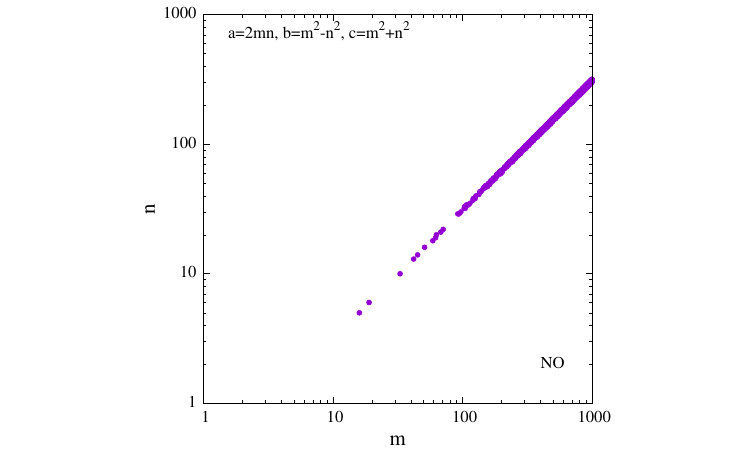}
\includegraphics[keepaspectratio, scale=0.7]{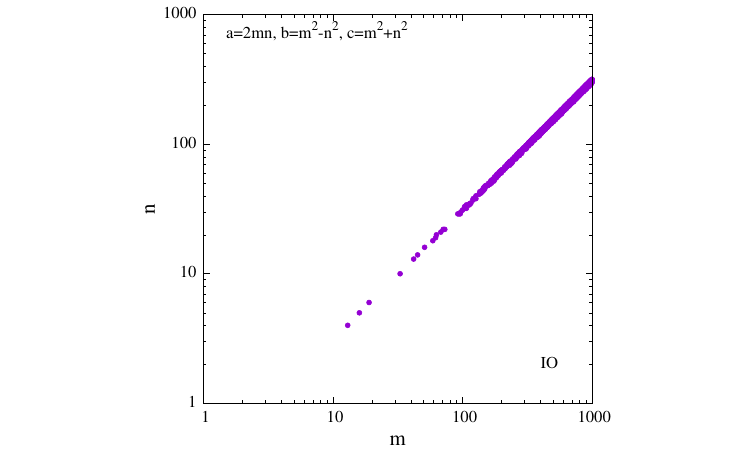}
 \caption{Similar to Fig. \ref{fig:m_n_PM1C1}, except for $\left(a,b,c\right)=\left(2mn, m^2-n^2,  m^2+n^2\right)$ and $\left(U_{\rm{PM}}\right)_{\rm{C2}}$.}
 \label{fig:m_n_PM2C2}
  \end{figure*}
%--------------------------------------------------------------------
%--------------------------------------------------------------------
\begin{figure*}[th]
\includegraphics[keepaspectratio, scale=0.7]{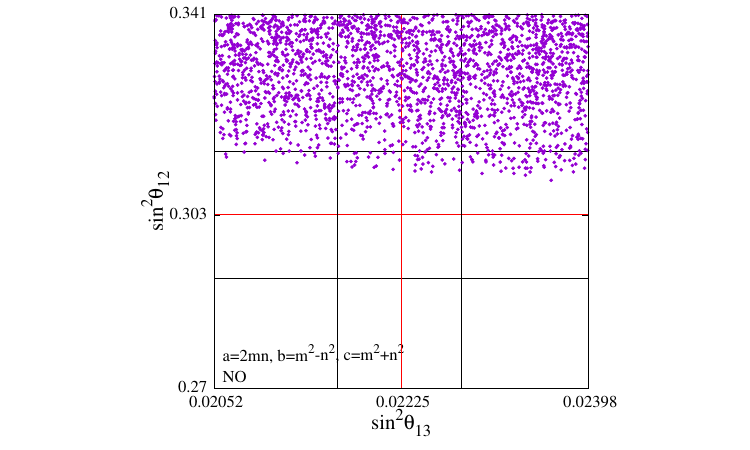}
\includegraphics[keepaspectratio, scale=0.7]{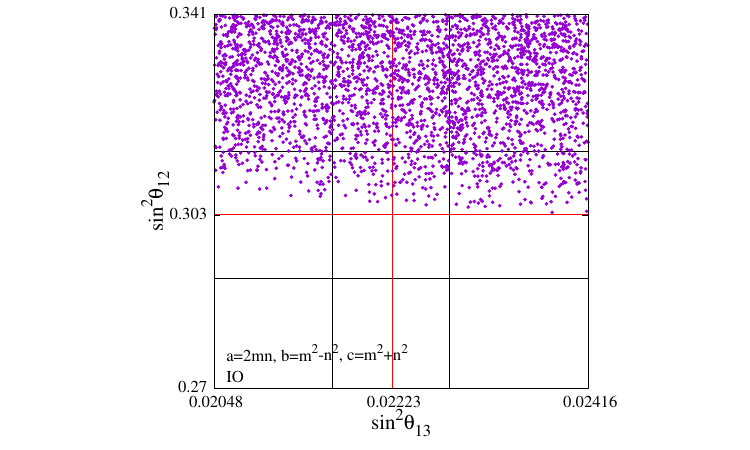}
\includegraphics[keepaspectratio, scale=0.7]{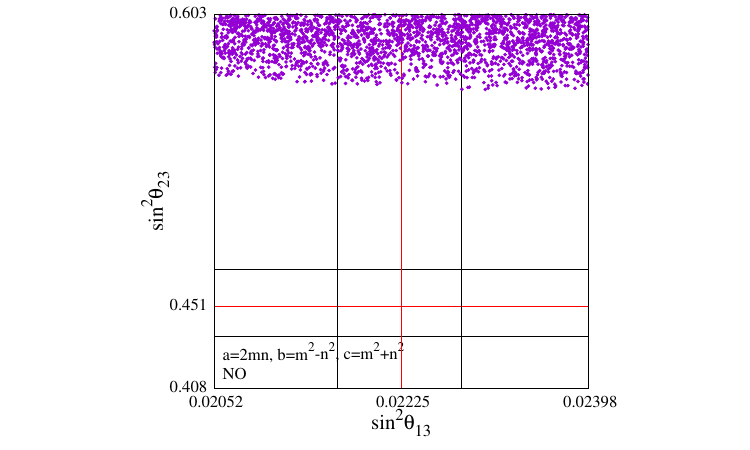}
\includegraphics[keepaspectratio, scale=0.7]{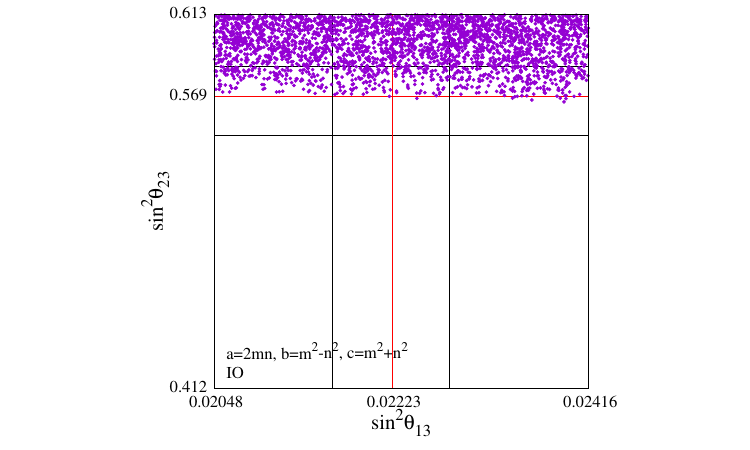}
 \caption{Similar to Fig. \ref{fig:12_13_23_PM1C1}, except for $\left(a,b,c\right)=\left(2mn, m^2-n^2,  m^2+n^2\right)$ and $\left(U_{\rm{PM}}\right)_{\rm{C2}}$.}
 \label{fig:12_13_23_PM2C2}
  \end{figure*}
%--------------------------------------------------------------------
%--------------------------------------------------------------------
\begin{figure*}[t]
\includegraphics[keepaspectratio, scale=0.7]{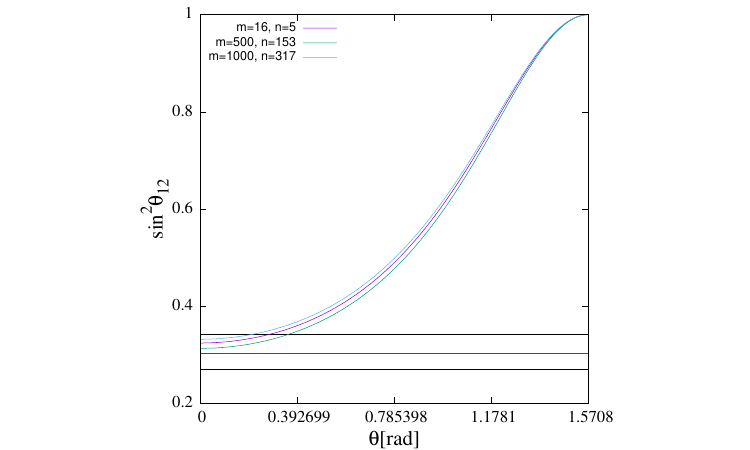}\\
\includegraphics[keepaspectratio, scale=0.7]{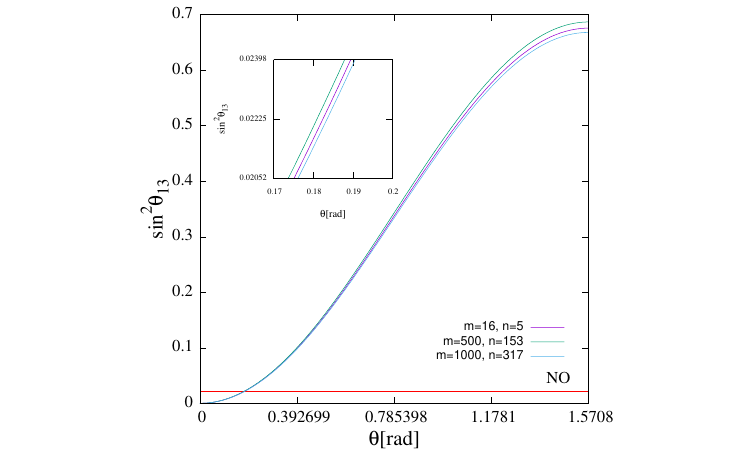}
\includegraphics[keepaspectratio, scale=0.7]{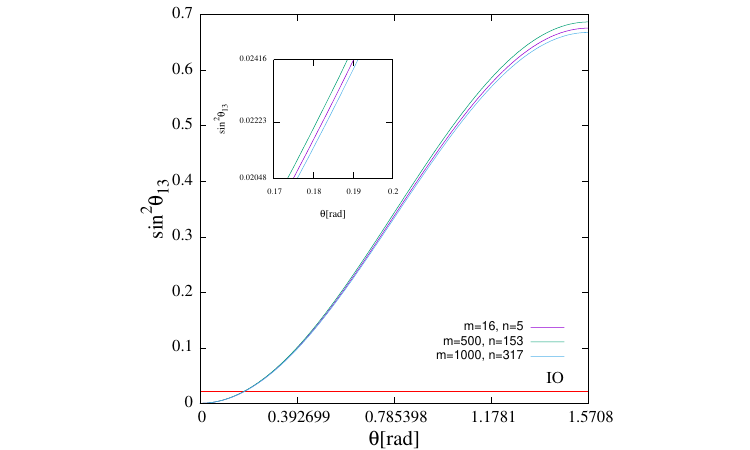}
 \caption{Similar to Fig. \ref{fig:12_13_t_PM1C1}, except for $\left(a,b,c\right)=\left(2mn, m^2-n^2,  m^2+n^2\right)$ and $\left(U_{\rm{PM}}\right)_{\rm{C2}}$.We selected $(m,n)=(16,5),(500,153),(1000,317)$.}
 \label{fig:12_13_t_PM2C2}
  \end{figure*}
%--------------------------------------------------------------------
%--------------------------------------------------------------------
\begin{figure*}[th]
\includegraphics[keepaspectratio, scale=0.7]{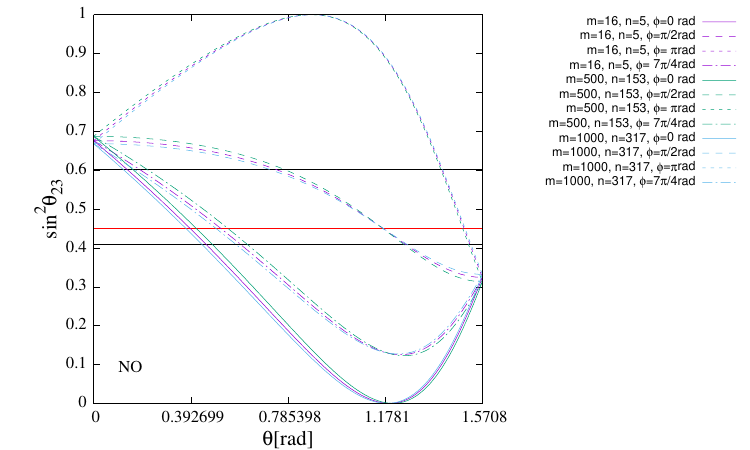}
\includegraphics[keepaspectratio, scale=0.7]{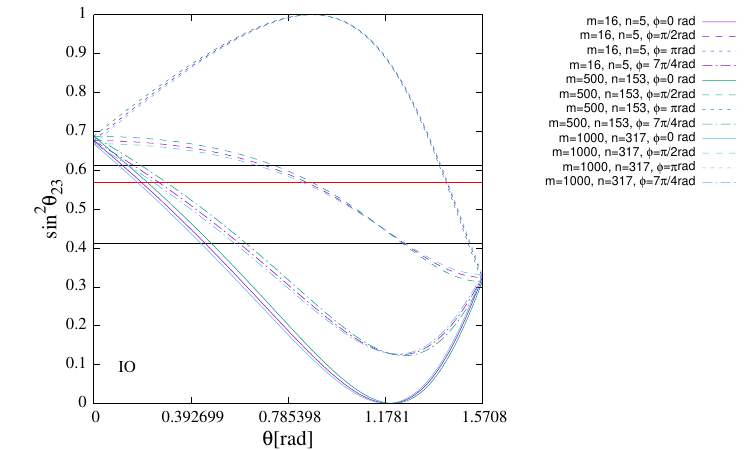}
 \caption{Similar to Fig. \ref{fig:23_t_PM1C1}, except for $\left(a,b,c\right)=\left(2mn, m^2-n^2,  m^2+n^2\right)$ and $\left(U_{\rm{PM}}\right)_{\rm{C2}}$. We selected $(m,n)=(16,5),(500,153),(1000,317)$ and $\phi=0, \frac{\pi}{2},\pi,\frac{7\pi}{4}$.}
 \label{fig:23_t_PM2C2}
  \end{figure*}
%--------------------------------------------------------------------
%--------------------------------------------------------------------
\begin{figure*}[th]
\includegraphics[keepaspectratio, scale=0.7]{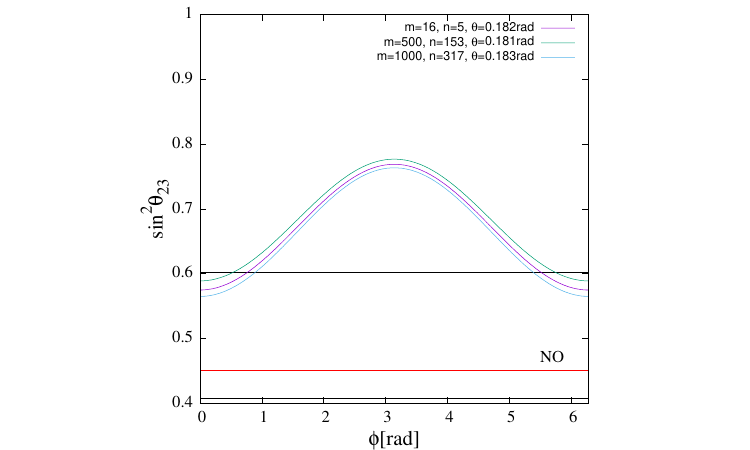}
\includegraphics[keepaspectratio, scale=0.7]{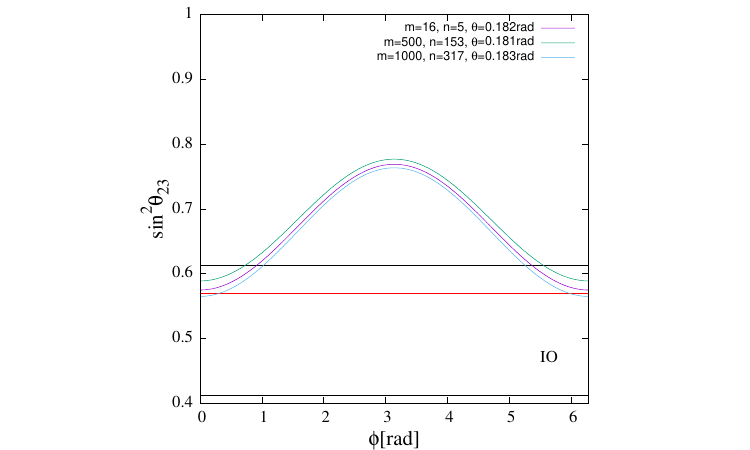}
 \caption{Similar to Fig. \ref{fig:23_p_PM1C1}, except for $\left(a,b,c\right)=\left(2mn, m^2-n^2,  m^2+n^2\right)$ and $\left(U_{\rm{PM}}\right)_{\rm{C2}}$. We selected $(m,n)=(16,5),(500,153),(1000,317)$.}
 \label{fig:23_p_PM2C2}
  \end{figure*}
%--------------------------------------------------------------------
%--------------------------------------------------------------------
\begin{figure*}[th]
\includegraphics[keepaspectratio, scale=0.7]{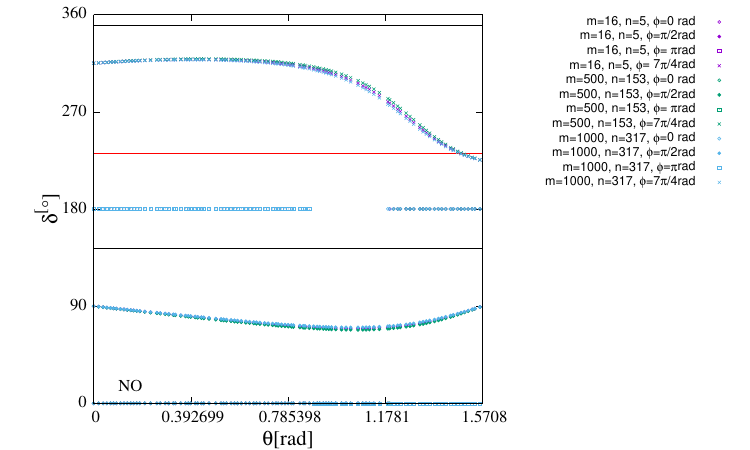}
\includegraphics[keepaspectratio, scale=0.7]{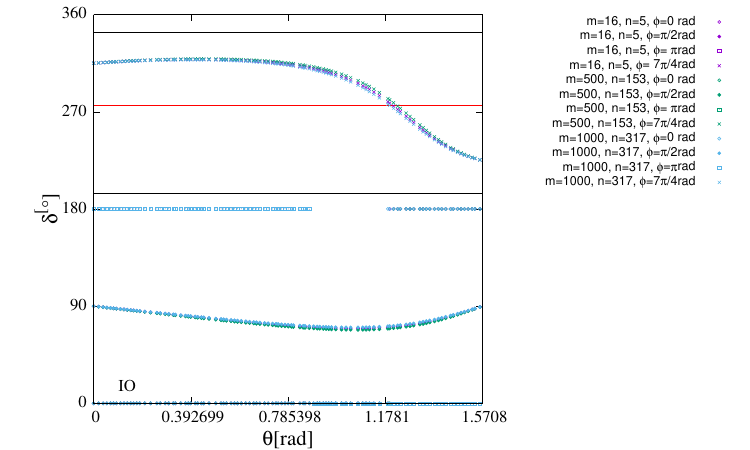}
 \caption{Similar to Fig. \ref{fig:d_t_PM1C1}, except for $\left(a,b,c\right)=\left(2mn, m^2-n^2,  m^2+n^2\right)$ and $\left(U_{\rm{PM}}\right)_{\rm{C2}}$. We selected $(m,n)=(16,5),(500,153),(1000,317)$ and $\phi=0, \frac{\pi}{2},\pi,\frac{7\pi}{4}$.}
 \label{fig:d_t_PM2C2}
  \end{figure*}
%--------------------------------------------------------------------
%--------------------------------------------------------------------
\begin{figure*}[th]
\includegraphics[keepaspectratio, scale=0.7]{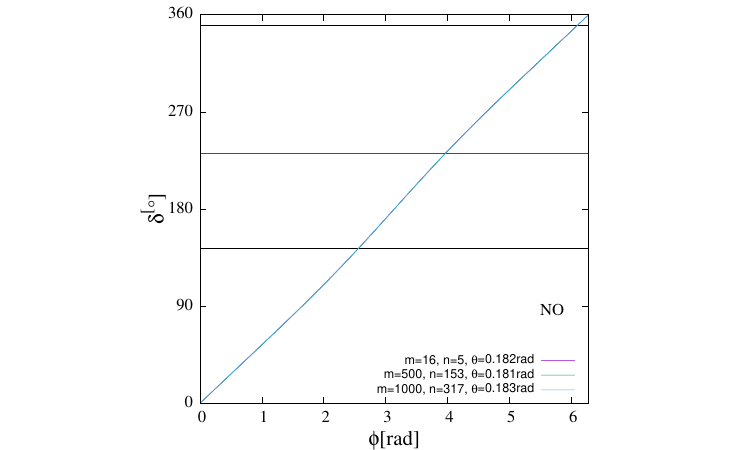}
\includegraphics[keepaspectratio, scale=0.7]{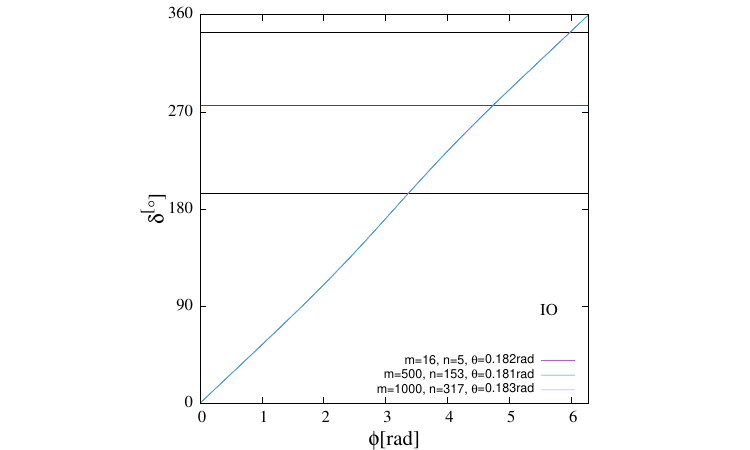}
 \caption{Similar to Fig. \ref{fig:d_p_PM1C1}, except for $\left(a,b,c\right)=\left(2mn, m^2-n^2,  m^2+n^2\right)$ and $\left(U_{\rm{PM}}\right)_{\rm{C2}}$. We selected $(m,n)=(16,5),(500,153),(1000,317)$.}
 \label{fig:d_p_PM2C2}
  \end{figure*}
%--------------------------------------------------------------------
Fig. \ref{fig:m_n_PM2C2} is the same as Fig. \ref{fig:m_n_PM1C1}, except for $\left(a,b,c\right)=\left(2mn, m^2-n^2,  m^2+n^2\right)$ and $\left(U_{\rm{PM}}\right)_{\rm{C2}}$. The key observations from Fig. \ref{fig:m_n_PM2C2} are as follows:
\begin{itemize}
   \item In the NO and IO cases, the PPMCs were both $0.999$. The PPMCs were close to $1$; thus, a very strong positive linear correlation was observed between $m$ and $n$.
 \item The relationship between $m$ and $n$ exhibits strong linearity. Thus, using Eq. (\ref{eq:lineari}), $w$ was $0.311$ and $0.310$ in the NO and IO cases, respectively. If $m$ increased by $10$, $n$ increased by $3$. $\beta$ was $-0.0995$ and $-0.0370$ in the NO and IO cases, respectively. $w$ and $\beta$ did not significantly vary between the NO and IO cases.
  \item The slop of Fig. \ref{fig:m_n_PM2C2} is smaller than that of Fig. \ref{fig:m_n_PM1C2}.
\end{itemize}

Fig. \ref{fig:12_13_23_PM2C2} is the same as Fig. \ref{fig:12_13_23_PM1C1}, except for $\left(a,b,c\right)=\left(m^2-n^2, 2mn, m^2+n^2\right)$ and $\left(U_{\rm{PM}}\right)_{\rm{C2}}$. The key observations from Fig. \ref{fig:12_13_23_PM2C2} are as follows:
\begin{itemize}
 \item In the NO and IO cases, the predicted values of $s_{12}^2$, $s_{13}^2$, and $s_{23}^2$ from the model fell within the $3 \sigma$ region.
 \item The predicted values from the model exceeded or were equal to $0.3$ for $s_{12}^2$ and $0.55$ for $s_{23}^2$ in both case. Thus, it was impossible to simultaneously satisfy the best-fit values for $s_{12}^2$, $s_{13}^2$, and $s_{23}^2$ in the case of NO.
\end{itemize}

A benchmark point,
\begin{eqnarray}
(m, n, \theta, \phi) = (16,5,0.1881~\rm{rad}, 5.529~ \rm{rad})
\label{Eq:PM1C1_p}
\end{eqnarray}
yields
\begin{eqnarray}
(s_{12}^2,s_{13}^2,s_{23}^2,\delta) = (0.3321, 0.02363, 0.5986, 319.2^\circ).
\label{Eq:PM1C1_mixing_angle}
\end{eqnarray}

Next, using $(m,n)=(16,5),(500,153),(1000,317)$, we investigated the relationship between the neutrino mixing angles, $\delta$, rotation angle $\theta$, and phase parameter $\phi$.

Fig. \ref{fig:12_13_t_PM2C2} is the same as  Fig. \ref{fig:12_13_t_PM2C1}, except for $\left(a,b,c\right)=\left(2mn, m^2-n^2,  m^2+n^2\right)$ and $\left(U_{\rm{PM}}\right)_{\rm{C2}}$. The key observations from Fig. \ref{fig:12_13_t_PM2C2} are as follows: 
\begin{itemize}
\item $\theta$ exists to satisfy the 3$\sigma$ region of $s_{13}^2$ and $s_{12}^2$.
 \item The range of $\theta$ satisfying the 3$\sigma$ region of $s_{13}^2$ was limited to $0.170~\rm{rad}\leq\theta\leq0.190~\rm{rad}$.
 \item If the range of $\theta$ was $0.170~\rm{rad}\leq\theta\leq0.190~\rm{rad}$, $s_{12}^2$ was within the 3$\sigma$ region of $s_{12}^2$. However, it exceeded the best-fit value.
\end{itemize}

Fig. \ref{fig:23_t_PM2C2} is the same as Fig. \ref{fig:23_t_PM1C1}, except for $\left(a,b,c\right)=\left(2mn, m^2-n^2,  m^2+n^2\right)$ and $\left(U_{\rm{PM}}\right)_{\rm{C2}}$. Fig. \ref{fig:23_p_PM2C2} is the same as  Fig. \ref{fig:23_p_PM1C1}, except for $\left(a,b,c\right)=\left(2mn, m^2-n^2,  m^2+n^2\right)$ and $\left(U_{\rm{PM}}\right)_{\rm{C1}}$. The key observations from Fig. \ref{fig:23_t_PM2C2} and Fig. \ref{fig:23_p_PM2C2} are as follows:

\begin{itemize}
 \item $\theta$ and $\phi$ exist to satisfy the 3$\sigma$ region of $s_{23}^2$.
 \item This combination of $m$ and $n$ satisfies the best-fit value for IO.
 \item This combination of $m$ and $n$ does not satisfy the best-fit value for NO.
 \item If $s_{13}^2$ aligns with the best-fit value, $s_{23}^2$ was close to the upper limit of the 3$\sigma$ region.
  \item As shown in Fig. \ref{fig:23_p_PM2C2}, the range of $\phi$ satisfying the 3$\sigma$ region of $s_{23}^2$ was limited to $0~\rm{rad} \leq\phi\leq1.0~\rm{rad}$ and $5.3~\rm{rad} \leq\phi\leq2\pi~\rm{rad}$.
\end{itemize}

Fig. \ref{fig:d_t_PM2C2} is the same as Fig. \ref{fig:d_t_PM1C1}, except for $\left(a,b,c\right)=\left(2mn, m^2-n^2,  m^2+n^2\right)$ and $\left(U_{\rm{PM}}\right)_{\rm{C2}}$. Fig. \ref{fig:d_p_PM2C2} is the same as Fig. \ref{fig:d_p_PM1C1}, except for $\left(a,b,c\right)=\left(2mn, m^2-n^2,  m^2+n^2\right)$ and $\left(U_{\rm{PM}}\right)_{\rm{C2}}$. The key observations from Figs.\ref{fig:d_t_PM2C2} and \ref{fig:d_p_PM2C2} are as follows:
\begin{itemize}
 \item No difference was observed in the $\delta$ for different combinations of $m$ and $n$.
 \item As shown in Fig. \ref{fig:d_p_PM2C2}, the range of $\phi$ satisfying the 3$\sigma$ region of $\delta$ was limited to $2.6~\rm{rad} \leq\phi\leq6.0~\rm{rad}$.
\end{itemize}
%=================================================
%=================================================
\section{$Z_2$ Symmetry\label{section:Z2}}
%%----------------------------------------------------------------------------------
The topic of symmetry is crucial in neutrino physics. As an example, a texture of the neutrino mass matrix based on the $A_n$ or $S_n$ symmetry has been proposed Ref. \cite{Altarelli2010RMP}. Discussions regarding the $Z_2$ symmetry are conducted in the context of TBM or TM mixing models \cite{LamPRD2006,LamPLB2007,LamPRL2008}. Ref. \cite{DicusPRD2011} discusses the $Z_2$ symmetry of the mixing matrix proposed in Ref. \cite{DicusPRD2011}. Our PM exhibits $Z_2$ symmetry. Herein, we discuss the $Z_2$ symmetry of PM.

Assuming that the mass matrix of charged leptons is diagonal and real, the neutrino mass matrix, $M_{\nu}$, is as follows:
\begin{eqnarray}
M_{\nu}=U^{\ast}\textrm{diag} (m_1,m_2,m_3)U^{\dag}
=\left(
\begin{array}{ccc}
A&B&C\\
B&D&E\\
C&E&F
\end{array}
\right),
\label{Eq:mass_matrix}
\end{eqnarray}
where $m_1$, $m_2$, and $m_3$ are the neutrino mass eigenvalues.

If $M_{\nu}$ satisfies the transformation,
\begin{eqnarray}
G^T M_{\nu} G&=&M_{\nu},\nonumber\\
G^2&=&\textrm{diag}(1,1,1).
\label{Eq:Z2}
\end{eqnarray}
$M_{\nu}$ is invariant under $Z_2$ symmetry\cite{LamPRD2006}. In this context, $G$ is expressed as follows: 
\begin{eqnarray}
G&=&g_1v_1v_1^{\dag}+g_2v_2v_2^{\dag}+g_3v_3v_3^{\dag},
\label{Eq:G_Z2}
\end{eqnarray}
where $v_1$, $v_2$, and $v_3$ are column vectors of the mixing matrix, and one of the eigenvalues among $g_1$, $g_2$, and $g_3$ was equal to $-1~(1)$. The remaining two eigenvalues were equal to $1~(-1)$\cite{LamPRD2006,LamPLB2007,LamPRL2008}.

The column vectors of the PM are
\begin{eqnarray}
v_1&=&\left(\frac{b}{c},-\frac{a^2}{c^2},\frac{ab}{c^2}\right)^T,\nonumber\\
v_2&=&\left(\frac{a}{c},-\frac{ab}{c^2},-\frac{b^2}{c^2}\right)^T,\nonumber\\
v_3&=&\left(0,\frac{b}{c},\frac{a}{c}\right)^T.
\label{Eq:PM_Z2_CV}
\end{eqnarray}
Based on Eq. (\ref{Eq:G_Z2}), $G_i~(i=1,2,3)$ can be expressed as
\begin{eqnarray}
&&G_1=-v_1\left(v_1\right)^{\dag}+v_2\left(v_2\right)^{\dag}+v_3\left(v_3\right)^{\dag}\nonumber \\
&&=\left(
\begin{array}{ccc}
\frac{(a-b)(a+b)}{c^2}&\frac{2a^2b}{c^3}&-\frac{2ab^2}{c^3}\\
\frac{2a^2b}{c^3}&\frac{-a^4+a^2b^2+b^2c^2}{c^4}&\frac{ab(a^2-b^2+c^2)}{c^4}\\
-\frac{2ab^2}{c^3}&\frac{ab(a^2-b^2+c^2)}{c^4}&\frac{b^4+a^2(c-b)(c+b)}{c^4}
\end{array}
\right), 
\label{Eq:PM_G1}
\end{eqnarray}
\begin{eqnarray}
&&G_2=v_1\left(v_1\right)^{\dag}-v_2\left(v_2\right)^{\dag}+v_3\left(v_3\right)^{\dag}\nonumber \\
&&=\left(
\begin{array}{ccc}
\frac{(b+a)(b-a)}{c^2}&-\frac{2a^2b}{c^3}&\frac{2ab^2}{c^3}\\
-\frac{2a^2b}{c^3}&\frac{a^4-a^2b^2+b^2c^2}{c^4}&\frac{ab(-a^2+b^2+c^2)}{c^4}\\
\frac{2ab^2}{c^3}&\frac{ab(-a^2+b^2+c^2)}{c^4}&\frac{-b^4+a^2(b^2+c^2)}{c^4}
\end{array}
\right),
\label{Eq:PM_G2}
\end{eqnarray}
and,
\begin{eqnarray}
&&G_3=v_1\left(v_1\right)^{\dag}+v_2\left(v_2\right)^{\dag}-v_3\left(v_3\right)^{\dag}\nonumber \\
&&=\left(
\begin{array}{ccc}
\frac{a^2+b^2}{c^2}&0&0\\
0&\frac{a^4+a^2b^2-b^2c^2}{c^4}&-\frac{ab(a^2+b^2+c^2)}{c^4}\\
0&-\frac{ab(a^2+b^2+c^2)}{c^4}&\frac{b^4+a^2(b-c)(b+c)}{c^4}
\end{array}
\right),
\label{Eq:PM_G3}
\end{eqnarray}
where $G_1^2=G_2^2=G_3^2=\textrm{diag}(1,1,1)$.

The neutrino mass matrix obtained from $\left(U_{\rm{PM}}\right)_{\rm{C1}}$ is written as follows:
\begin{eqnarray}
M_{\left(U_{\rm{PM}}\right)_{\rm{C1}}}=\left(U_{\rm{PM}}\right)_{\rm{C1}}^{\ast}\textrm{diag} (m_1,m_2,m_3)\left(U_{\rm{PM}}\right)_{\rm{C1}}^{\dag}\nonumber
\end{eqnarray}
\begin{widetext}
\begin{eqnarray}
=\left(
\begin{array}{ccc}
A&B&C\\
B&D&\frac{a^4B-b^4B-a^3bC-a^2b\left\{bB+c\left(A-D\right)\right\}}{ab^2c}\\
C&\frac{a^4B-b^4B-a^3bC-a^2b\left\{bB+c\left(A-D\right)\right\}}{ab^2c}&\frac{a^5B+ab^3\left(cA-bB\right)-a^4bC-b^5C-a^3b\left\{2bB+c\left(A-D\right)\right\}}{ab^3c}
\end{array}
\right). 
\end{eqnarray}
\end{widetext}
The neutrino mass matrix obtained from $\left(U_{\rm{PM}}\right)_{\rm{C2}}$ is written as follows:
\begin{eqnarray}
M_{\left(U_{\rm{PM}}\right)_{\rm{C2}}}&=&\left(U_{\rm{PM}}\right)_{\rm{C2}}^{\ast}\textrm{diag} (m_1,m_2,m_3)\left(U_{\rm{PM}}\right)_{\rm{C2}}^{\dag}\nonumber
\end{eqnarray}
\begin{widetext}
\begin{eqnarray}
=\left(
\begin{array}{ccc}
A&B&C\\
B&D&\frac{a^3B+b^3C+abc\left(D-A\right)}{b^2c}\\
C&\frac{a^3B+b^3C+abc\left(D-A\right)}{b^2c}&\frac{a^5B+ab^3\left(cA+bB\right)+a^4bC+2a^2b^3C-b^5C+a^3bc\left(D-A\right)}{ab^3c}
\end{array}
\right).
\label{Eq:PMC2_mass_matrix}
\end{eqnarray}
\end{widetext}
Considering that these neutrino mass matrices, $M_{\left(U_{\rm{PM}}\right)_{\rm{C1}}}$ and $M_{\left(U_{\rm{PM}}\right)_{\rm{C2}}}$, were satisfied through transformation
\begin{eqnarray}
G_1^TM_{\left(U_{\rm{PM}}\right)_{\rm{C1}}}G_1=M_{\left(U_{\rm{PM}}\right)_{\rm{C1}}},
\end{eqnarray}
and 
\begin{eqnarray}
G_2^TM_{\left(U_{\rm{PM}}\right)_{\rm{C2}}}G_2=M_{\left(U_{\rm{PM}}\right)_{\rm{C2}}},
\end{eqnarray}
they were invariant under $Z_2$ symmetry.
%=================================================
%=================================================
\section{Summary\label{section:summary}}
%%----------------------------------------------------------------------------------
Here, we focused on primitive Pythagorean triples \cite{Williams2017,MartensarXiv2112,Kocik2007,SchmelzerarXiv2021,YekutieliarXiv2021,Cha2018,DeriyarXiv2023,PricearXiv2008,AlperinarXiv2000} to construct a neutrino mixing model. Primitive Pythagorean triples refer to sets of three natural numbers $(a, b, c)$ satisfying $c^2=a^2+b^2$. For instance, in the triangle, $(3, 4, 5)$, the three internal angles are $(36.87^\circ,53.13^\circ,90.00^\circ)$. Among these, $(36.87^\circ,53.13^\circ)$ closely approximate the upper limit of $3 \sigma$ for the solar neutrino mixing angle, $35.74^\circ$, and the upper limit of the $3 \sigma$ atmospheric mixing angle, $51.0^\circ~\left(51.5^\circ\right)$, in the case of NO (IO).

First, the relationship between primitive Pythagorean triples and neutrino mixing angles, $\theta_{12}$ and $\theta_{23}$, is as follows:
\begin{eqnarray} 
s_{12}=\frac{a}{c},~~s_{23}=\frac{b}{c},
\end{eqnarray}
where the relationship among $a$, $b$, and $c$ is described by $a<b<c$. From this relationship, we constructed the neutrino mixing matrix $U_{\rm{PM}}$. However, the mixing matrix predicts the vanishing reactor mixing angle. To improve the reproducibility of the reactor mixing angle, we modified $\left(U_{\rm{PM}}\right)_{\rm{C1}}=U_{\rm{PM}}U_{23}$ ($\left(U_{\rm{PM}}\right)_{\rm{C2}}=U_{\rm{PM}}U_{13}$) where $U_{23}$ and $U_{13}$ are
\begin{eqnarray}
U_{23}=\left(
\begin{array}{ccc}
1&0&0\\
0&\cos{\theta}&\sin{\theta}e^{-i\phi}\\
0&-\sin{\theta}e^{i\phi}&\cos{\theta}
\end{array}
\right)
\end{eqnarray}
and
\begin{eqnarray}
U_{13}=\left(
\begin{array}{ccc}
\cos{\theta}&0&\sin{\theta}e^{-i\phi}\\
0&1&0\\
-\sin{\theta}e^{i\phi}&0&cos{\theta}
\end{array}
\right).
\end{eqnarray}
Consequently, we obtained a new neutrino mixing model related to primitive Pythagorean triples that satisfies the observed values.

We showed that the neutrino mass matrix, $M_{\left(U_{\rm{PM}}\right)_{\rm{C1}}}$ $\left(M_{\left(U_{\rm{PM}}\right)_{\rm{C2}}} \right)$, obtained from $\left(U_{\rm{PM}}\right)_{\rm{C1}}$ $\left(\left(U_{\rm{PM}}\right)_{\rm{C2}}\right)$ was invariant under $Z_2$ symmetry.

Finally, in this paper, we modified $U_{\rm{PM}}$ into $U_{\rm{PM}}U_{23}$ ($U_{\rm{PM}}U_{13}$). We would like to consider the charged lepton mixing, $U_e$, as in Ref. \cite{PetcovNPB2015} and the texture of the neutrino mass matrix in the future. In the present numerical calculation results, the simplest primitive Pythagorean triple, $(3,4,5)$, did not match the observed values. The simplest primitive Pythagorean triple $(3,4,5)$ shared relations with other primitive Pythagorean triples, as observed in Ref. \cite{Cha2018,DeriyarXiv2023,PricearXiv2008,AlperinarXiv2000}. Such relationships are referred to as the trees of primitive Pythagorean triples. We would like to investigate the relationship between the trees of primitive Pythagorean triples and the neutrino mixing  in the future.
%=================================================
\bibliography{apssamp}% Produces the bibliography via BibTeX.

\begin{thebibliography}{0}    %for 1 digit
%\bibitem{Kajiyama2007}
%Y. Kajiyama, M. Raidal, and A. Strumia, \Journal{\PRD}{76}{117301}{2007}.
%%Pythagorean triple%%%%%
\bibitem{Williams2017}
K. R. Williams, TRIPLES: Applictions of Pythagorean Triples (Inspiration Books, 2017)
\bibitem{MartensarXiv2112}
G. J. Martens, arXiv:2112.09553 [math.GM].
\bibitem{Kocik2007}
J. Kocik,\Journal{\AACA}{17}{71-93}{2007}.
\bibitem{SchmelzerarXiv2021}
A. Schmelzer, and S. Chetty, arXiv:2102.03451 [math.NT].
\bibitem{YekutieliarXiv2021}
A. Yekutieli, arXiv:2101.12166 [math.NT].
\bibitem{Cha2018}
B. Cha, E. Nguuyen, and B. Tauber, \Journal{\JNP}{185}{218-256}{2018}.
\bibitem{DeriyarXiv2023}
P. Deriy, arXiv:2310.15174 [math.HO].
\bibitem{PricearXiv2008}
H. Lee. Price, arXiv:0809.4324 [math.HO].
\bibitem{AlperinarXiv2000}
R. C. Alperin, arXiv:0010281 [math.HO].
%%%%%PMNS%%%%%%%%%
\bibitem{Pontecorvo1957}
B. Pontecorvo,  \Journal{\SPJETP}{6}{429}{1957}.
\bibitem{Pontecorvo1958}
B. Pontecorvo,  \Journal{\SPJETP}{7}{172}{1958}.
\bibitem{Maki1962PTP}
Z. Maki, M. Nakagawa and S. Sakata, \Journal{\PTP}{28}{870}{1962}.
\bibitem{PDG}
M. Tanabashi {\it et al.} (Particle Data Group), \Journal{\PRD}{98}{030001}{2018}.
\bibitem{Jarlskog}
C. Jarlskog, \Journal{\PRL}{55}{1039}{1985}.
\bibitem{NuFit}
I. Esteban, M. C. Gonzalez-Garcia, M. Maltoni, T. Schwetz, and A. Zhou, \Journal{\JHEP}{09}{178}{2020}. See also, NuFIT 5.2 (2022), www.nu-fit.org.
\bibitem{Xing2007PLB}
Z. Z. Xing and S. Zhou, \Journal{\PLB}{653}{278}{2007}.
\bibitem{Harrison2006PRD}
J. D. Bjorken, P. F. Harrison, and W. G. Scott, \Journal{\PRD}{74}{073012}{2006}.
%%PPMCC%%%%
\bibitem{Pearson1895}
K. Pearson, Proc. R. Soc. Lond. {\bf58}, 240 (1895).
%%%An,Sn%%%
\bibitem{Altarelli2010RMP}
G. Altarelli and F. Feruglio, \Journal{\RMP}{82}{2701}{2010}.
 %%%%%Z2%%%%%%%
\bibitem{LamPRD2006}
C. S. Lam, \Journal{\PRD}{74}{113004}{2006}.
\bibitem{LamPLB2007}
C. S. Lam, \Journal{\PLB}{656}{193}{2007}.
\bibitem{LamPRL2008}
C. S. Lam, \Journal{\PRL}{101}{121602}{2008}.
\bibitem{DicusPRD2011}
D. A. Dicus, S. -F. Ge, and W. W. Repko, \Journal{\PRD}{83}{093007}{2011}.
%%%%
\bibitem{PetcovNPB2015}
S. T. Petcov, \Journal{\NPB}{892}{400-428}{2015}.
\end{thebibliography}

\end{document}